\documentclass[amsmath,twocolumn,superscriptaddress,nature,aps]{revtex4-2}
\usepackage{amsthm,amsfonts,graphicx,color,amsmath,amssymb,textcomp}
\usepackage{graphicx}
\usepackage{wasysym}
\usepackage{amsfonts}
\usepackage{bm}
\usepackage{enumerate}
\usepackage{color}
\usepackage{epstopdf}
\usepackage{latexsym}
\usepackage[breaklinks,colorlinks = true,linkcolor = red,urlcolor=cyan,citecolor=red]{hyperref}
\usepackage[caption=false,singlelinecheck=false]{subfig}

\usepackage[compat=1.1.0]{tikz-feynman}
\usepackage{times}
\usepackage{float}
\usepackage{changes}
\colorlet{Changes@Color}{black}
\usepackage{gensymb}
\usepackage{comment}

\usepackage{times}
\newcommand{\bea}{\begin{eqnarray}}
\newcommand{\eea}{\end{eqnarray}}
\newcommand{\be}{\begin{eqnarray}}
\newcommand{\ee}{\end{eqnarray}}
\newcommand{\bw}{\begin{widetext}}
\newcommand{\ew}{\end{widetext}}
\newcommand{\nn}{\nonumber}

\newcommand{\la}{\langle}
\newcommand{\ra}{\rangle}

\newcommand{\tbf}{\textbf}

\begin{document}
\title{Displacement-Field-Driven Transition between Superconductivity and Valley Ferromagnetism in Transition Metal Dichalcogenides}

\author{Hyeok-Jun Yang}
\email{hyang23@nd.edu}
\affiliation{Department of Physics, University of Notre Dame, Notre Dame, Indiana 46556, USA}

\author{Yi-Ting Hsu}
\email{yhsu2@nd.edu}
\affiliation{Department of Physics, University of Notre Dame, Notre Dame, Indiana 46556, USA}

\date{\today}

\begin{abstract}
Recent experiments have observed transitions between superconductivity and correlated magnetism in twisted bilayer WSe$_2$ near van-Hove fillings, driven by the displacement field $D$. 
Motivated by the experiment, we theoretically propose a general mechanism for a $D$-controlled transition between  superconductivity and ferromagnetism in two-dimensional (2D) spin-orbit-coupled hexagonal systems, where van Hove singularities (VHS) lie on the Fermi level. 
We show that such a transition can be naturally captured by a simple VHS-only model without Fermi surface details, where the inter-VHS interactions that govern the Fermi surface instabilities  is controlled by $D$ through the band projection of screened Coulomb interaction.  
By treating this simple model with renormalization group technique beyond mean-field level, we find that a chiral $d/p$-wave superconductivity naturally dominates under a weak displacement field $D<D_c$. 
At a stronger displacement field $D>D_c$, a \textit{valley ferromagnetic phase} (vFM) takes over, which is spatially non-uniform due to valley-modulated magnetization. Finally, we discuss generic conditions for the predicted superconductivity-to-ferromagnetism transition to take place in the rich family of few-layer hexagonal van der Waals material systems. 
Taking twisted bilayer WSe$_2$ as a case study, we discuss experimental detections that can falsify our prediction.  
\end{abstract}
\maketitle


\textit{Introduction---} 
Two-dimensional superconductivity (SC) and its transitions into other symmetry-broken phases have recently been observed in a variety of van der Waals materials, tunable by external fields \cite{doi:10.1073/pnas.1620140114, PhysRevB.95.075420, PhysRevLett.121.026402, PhysRevLett.122.016401, Chen2019, Lu2019, Chen2019_2, Chen_2020, Tang2020, Cao2020, Kazmierczak2021}, carrier filling \cite{Cao2018_1, Cao2018_2, doi:10.1126/science.aav1910, Kerelsky2019, Choi2019_1,
doi:10.1126/science.aay5533,
doi:10.1126/science.aaw3780, PhysRevLett.123.046601, Wang2020, Xu2020,Jin2021}, and other experimental parameters \cite{PhysRevB.98.085144, Zhang2020}. Besides graphene-based systems, robust SC has recently been demonstrated in moiré transition metal dichalcogenides (TMDs) with strong spin-valley locking 
\cite{PhysRevLett.108.196802}. In particular, SC was seen in twisted bilayer WSe$_2$ (tWSe$_2$) at different twist angles, where transitions were observed from superconducting to other correlated phases upon increasing an external displacement field \cite{Xia2025, Guo2025}. 
The natures of the superconducting state   \cite{PhysRevB.104.195134, PhysRevB.108.064506, PhysRevB.108.155111, PhysRevLett.130.126001,PhysRevB.110.064516,PhysRevB.111.L060501,PhysRevLett.134.136503,guerci2024topologicalsuperconductivityrepulsiveinteractions,PhysRevB.110.035143, qin2024kohnluttingermechanismsuperconductivitytwisted} and the nearby correlated phases \cite{PhysRevB.111.014507, christos2024approximatesymmetriesinsulatorssuperconductivity, Kim2025} have been extensively studied, where the latter were widely believed to exhibit magnetism. For instance, 
recent 
magnetic dichroism measurement has found a ferromagnetic phase close to the Van Hove (VH) filling in the $\theta=3.2^\circ$ system \cite{Knüppel2025}, along with theories suggesting close competition among ferro- and antiferromagnetic phases \cite{peng2025magnetismtwistedbilayerwse2}. 
On the other hand, 
antiferromagnetic orders were proposed by mean-field \cite{tuo2024theorytopologicalsuperconductivityantiferromagnetic} and functional renormalization group (RG) \cite{fischer2024theoryintervalleycoherentafmorder} theories for the  $\theta=5^\circ$ system, while definitive experimental evidence for the magnetic order remains to be reported. 
The rapid experimental development and the rich variety of few-layer TMD systems thus call for generic mechanisms for SC-to-magnetism transitions that do not rely on system details.

\begin{figure}[t!]
{
\includegraphics[width=0.5\textwidth]{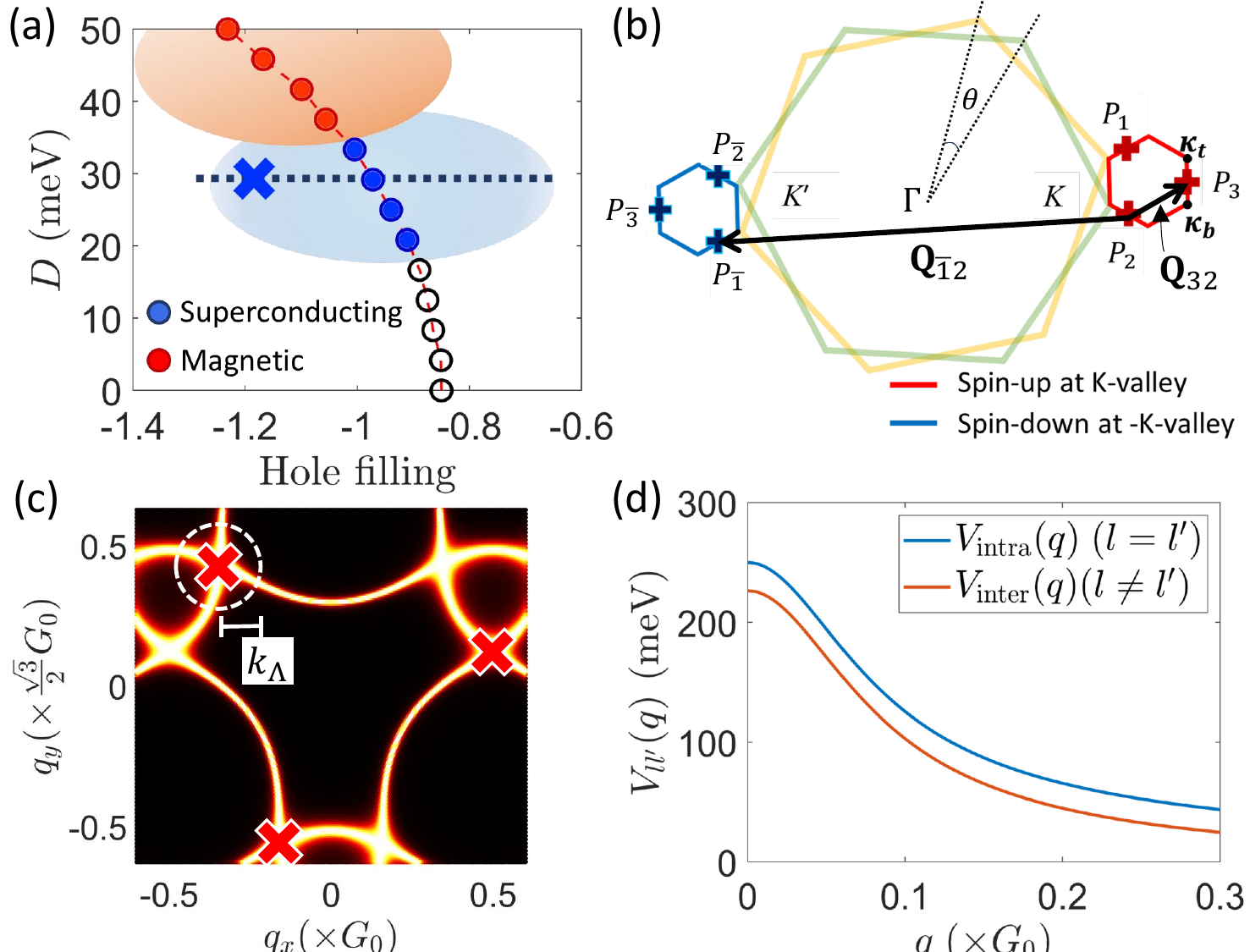}}
\caption{ 
(a) The phase diagram of tTMD as a function of hole filling $n$ and displacement field $D$. 
The red dashed line denotes the VH-filling curve $n_V(D)$ obtained from Eq. \ref{eq:Moire_H}. The filled circles label our RG results in Fig. \ref{fig:sus_plot}b, and the black dotted line labels where the mean-field calculation in Fig. \ref{fig:sus_plot}c is performed. The superconducting and magnetic phase regions away from $n=n_V(D)$ are shown schematically in blue and orange, respectively. 
The blue cross marks a representative point in the SC phase, where we produce Fig. \ref{fig:plot}b. 
(b) The spin-up and -down mBZ of a general tTMD at twist angle $\theta$, where $\tbf{P}_{1,2,3}$ and ${\tbf{P}}_{\bar{1},\bar{2},\bar{3}}=-\tbf{P}_{1,2,3}$ denote the VH-points at valley $K$ and $K'$, respectively. 
(c) The Fermi surface at valley $K$ at VH-filling for $D=30\text{meV}$, where the VH points $\tbf{P}_{1,2,3}$ are marked in red and 
the white dash line denotes the patch with size $k_{\Lambda}$ centered at each VH-point. 
(d) The momentum dependence of intra- and inter-layer Coulomb interactions $V_{\text{intra}}(q)\equiv V_{bb}(q)=V_{tt}(q)$ and $V_{\text{inter}}(q)\equiv V_{bt}(q)=V_{tb}(q)$ in Eq. \ref{eq:Coulomb}, where we set the layer separation $h=3a_0$, the moiré reciprocal vector $G_0=\frac{4\pi\theta}{\sqrt{3}a_0}$, the dielectric constant $\epsilon=25$, and the distance
to the gates $d =$ 20 nm. 
For the numerical plots in this work, we adapt the parameters for $\theta=5^\circ$ tWSe$_2$ \cite{Devakul2021}, where the effective mass $m^* = 0.45m_e$, the monolayer lattice constant $a_0=3.317$\r{A}, and
$(v,\psi, w)=(9.0\text{meV}, 128^{\circ}, 18\text{meV})$.}
\label{fig:schematic}
\end{figure}

In this work, we propose a mechanism for a displacement-field $D$-driven transition between SC and ferromagnetism in
two-dimensional spin-orbit-coupled hexagonal systems along the Van Hove filling curve $n_{V}(D)$ (see Fig. \ref{fig:schematic}a). 
In general, such a system can exhibit three VH singularities per valley on the Fermi level with appropriate gating  (see Fig. \ref{fig:schematic}b,c), where these six VH momenta evolve on the Fermi surface (FS) with $D$. 
For instance, in $\theta=5^\circ$ tWSe$_2$, the theoretically determined VH filling $n_{V}(D)$ was found to trace through the observed SC phase and the nearby magnetic phase, which agrees well with the Hall resistivity measurement  \cite{Guo2025}. 
Due to the diverging density of states at the VH momenta, a SC-to-magnetism transition in such spin-orbit-coupled hexagonal systems, if exists, should be driven by the evolution of these VHS under $D$, regardless of the Fermi surface details. 

Motivated by this hypothesis, within a simple model of six-VHS, we examine how the symmetry-allowed inter-VHS interactions receive a displacement field $D$-dependence from screened Coulomb interactions and renormalize in the long-wavelength limit. We show that among all possible instabilities, the renormalized inter-VHS interactions drive a mixed $p$- and $d$-wave chiral SC at $D<D_c$ and \textit{spatially non-uniform} valley ferromagnetism at $D>D_c$, where the displacement field $D$ triggers a Stoner-like transition. We expect that such a transition can generically occur in the family of few-layer TMD systems and other two-dimensional (2D) hexagonal spin-orbit-coupled systems
near the VH filling, in the weak to intermediate coupling regime.

\textit{Microscopic model---}
We start from a well-known continuum model for general twisted homobilayer TMDs, where the topmost valence bands in the Moire Brillouin zones (mBZ) at valley $K$ and $K$' are spin-up and -down, respectively, due to the Ising spin-orbit coupling (see Fig. \ref{fig:schematic}b).
The non-interacting model reads $H_0 = \sum_{\sigma =\uparrow, \downarrow}\int d^2\tbf{r} \psi_{\sigma}^\dagger(\tbf{r})  h_{\sigma}(\tbf{r}) \psi_{\sigma}(\tbf{r})$ \cite{PhysRevLett.122.086402}, where  $\psi_{\sigma}^\dagger(\tbf{r})$ creates an electron with spin $\sigma$ at spatial position $\tbf{r}$, and the spin-up Hamiltonian at valley $K$ is given by 
\bea
h_{\uparrow}(\tbf{r})
=
\begin{pmatrix}
-\frac{\hbar^2(\tbf{k}-\boldsymbol{\kappa}_t)^2}{2m^*}+\Delta_t(\tbf{r})
&
\Delta_{\text{T}}(\tbf{r}) \\
\Delta_{\text{T}}^\dagger(\tbf{r}) &
-\frac{\hbar^2(\tbf{k}-\boldsymbol{\kappa}_b)^2}{2m^*}+\Delta_b(\tbf{r})
\end{pmatrix},\quad\;
\label{eq:Moire_H}
\eea
where $h_{\downarrow}(\tbf{r})$ at valley $K'$ is related to Eq. \ref{eq:Moire_H} by time-reversal symmetry. 
In Eq. \ref{eq:Moire_H}, \tbf{k} is the in-plane crystalline momentum measured from $\Gamma$-point in Fig. \ref{fig:schematic}b, $l = t,b$ denotes the top and bottom layer, 
while $\Delta_{t,b}(\tbf{r}) = \pm \frac{eV_z}{2}+2v\sum_{j=1,3,5}\cos(\tbf{g}_j\cdot \tbf{r}\pm \psi)$ and $\Delta_{\text{T}}(\tbf{r}) = w(1+ e^{-i\tbf{g}_2\cdot \tbf{r}}+e^{-i\tbf{g}_3\cdot \tbf{r}})$ describe the layer-dependent potentials and inter-layer tunneling, respectively.  
The potential difference, $V_z$ is directly proportional to the displacement field $D$ through $V_z = D\times d/\epsilon$ where $d$ is the layer separation and $\epsilon$ is the dielectric constant. 
Here,  $\tbf{g}_{j=1,2,3}$ are the reciprocal lattice vectors of mBZ that can be obtained by rotating $\tbf{g}_1=\frac{4\pi \theta}{\sqrt{3}a_0}\hat{x}$ counterclockwise by an angle of $(j-1)\pi/3$, $j=2,3$.  
The resulting electronic dispersion generally exhibits three VHS at momenta $\tbf{P}_{1,2,3}$ between the mBZ corners $\boldsymbol{\kappa}_{t,b}$ in valley $K$, and another three at $\tbf{P}_{\bar{1},\bar{2},\bar{3}}=-\tbf{P}_{1,2,3}$ in valley $K'$ \cite{PhysRevB.104.195134} (see Fig. \ref{fig:schematic}c). 
When the electronic density $n$ is tuned to the Van Hove filling $n_{V}(D)$, at which the VH points sit on the FS, the density of state (DOS) becomes logarithmically divergent so that the FS becomes susceptible to a wide variety of instabilities under even weak-to-intermediate interactions.

We consider intra- and inter-layer screened Coulomb interactions projected onto the topmost band 
\bea
H_{\text{int}}
=\frac{1}{2A}
\sum_{ll'=t,b}\sum_{\tbf{k}}V_{ll'}(k) \rho_{l}(\tbf{k})
\rho_{l'}(-\tbf{k}), 
\label{eq:Coulomb}
\eea
where $A$ is the system area. In Eq. \ref{eq:Coulomb},  $V_{ll}(k)$ and $V_{l\bar{l}}(k)$ are the intra- and inter-layer Coulomb potentials at momentum $\tbf{k} =\tbf{q} + \tbf{G}$, respectively, where $\tbf{q}$ 
lies in the first mBZ, and $\tbf{G}$ is the moire reciprocal lattice vector
\footnote{By solving the Poisson equation with double-gate boundary conditions, we obtain 
the $k$-dependence of $V_{ll'}(k)$ in Fig. \ref{fig:schematic}d (see Supplementary Material (SM) Section I \cite{SM_tTMD}). Since $V_{ll'}(k)$ decays fast with $k$, we will neglect the $G\neq0$ components for simplicity.}.
Next, the density operator for layer $l$ can be expressed as $\rho_l=\sum_{\tau}\rho_{l,\tau}$ in terms of the  
band $\tau$-resolved density 
\bea
\rho_{l,\tau}(\tbf{q})=
\sum_{s=\uparrow, \downarrow}\sum_{\tbf{p}\in \text{mBZ}}
\Lambda_{s l \tau}(\tbf{p},\tbf{p}+\tbf{q})
c^{\dagger}_{s \tau}(\tbf{p})
c_{s \tau}(\tbf{p}+\tbf{q}),      \;\;
\label{eq:density_operator}
\eea 
where we will project $\rho_{l,\tau}$ onto the topmost band and omit the band label $\tau$ in the following.  
Here,   
$c_{s}(\tbf{p})$ annihilates spin-$s$ electrons at momentum $\tbf{p}$, and 
$\Lambda_{s l }(\tbf{p},\tbf{p}+\tbf{q})
=\la u_{s l}(\tbf{p})|u_{s l}(\tbf{p}+\tbf{q})\ra$
is given by the overlap between Bloch wavefunctions, which satisfies 
the normalization condition  $\sum_{l=t,b} \Lambda_{s l }(\tbf{p},\tbf{p}) =1$ for all $\tbf{p}$ and $s$. 
Note that although the Coulomb potentials $V_{ll'}$ does not depend on the displacement field $D$, the interaction $H_{\text{int}}$ in Eq. \ref{eq:Coulomb} still receives an explicit $D$-dependence through the wavefunction overlap $\Lambda_{s l}(D)$, leading to a $D$-driven phase transition. 

\begin{figure}[t!]
{
\includegraphics[width=0.5\textwidth]{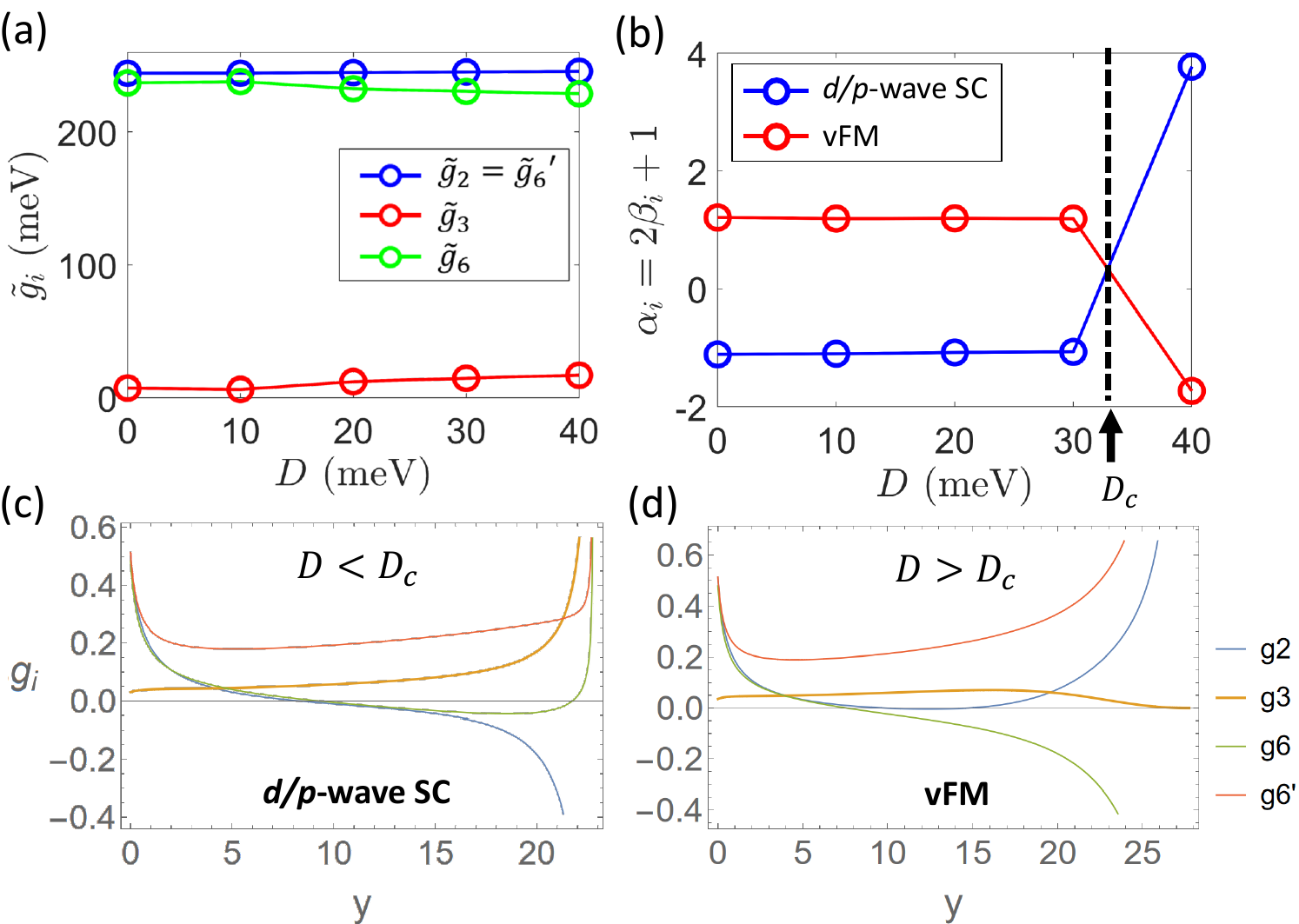}}
\caption{
(a) The bare interaction values of Eq. \ref{eq:Hg_i}, $\tilde{g}_i$ given by Eq. \ref{eq:g_and_V} which evolve as functions of $D$. 
(b) RG phase diagram along VH-filling. $d/p\text{-wave SC}$ phase is dominant for small $D<D_c$ while vFM phase is dominant for $D>D_c$ above the critical value $D_c \approx 33\text{meV}$. 
(c) RG flows evaluated at $D = 30$meV (in the $d/p$-wave superconducting phase) and (d) $D = 40$meV (in the ferromagnetic phase), highlighting qualitatively distinct behaviors across the phase boundary.
}
\label{fig:RG_plot}
\end{figure}

\textit{Patch model ---}
To investigate the FS instabilities at VH-fillings $n_{V}(D)$ at different displacement field strengths $D$, we first simplify the model $H=H_0+H_{\text{int}}$ by restricting the momentum $\textbf{k}$ to only the six  patches centered at VH points with sizes $k_\Lambda\ll G_0$ (see Fig. \ref{fig:schematic}c), rather than the full mBZs. This patch approximation is justified by the diverging DOS at the VH points. 
For the low-energy electrons lying within the patches, the screened Coulomb interactions $H_{\text{int}}$ in Eq. \ref{eq:Coulomb} now simplifies to  only four inequivalent inter-patch interactions due to the spin-valley locking \cite{PhysRevB.104.195134}
\bea
&&
H_\text{patch}
=
\sum_{n=1}^3
\Big[\tilde{g}_2 c^{\dagger}_{n}c_{n}c^{\dagger}_{\bar{n}}c_{\bar{n}}+\sum_{m\neq n}\Big(
\tilde{g}_3c^{\dagger}_{m}c_{n}c^{\dagger}_{\bar{m}}c_{\bar{n}}
\nn\\
&& \quad
+\frac{1}{2}\tilde{g}_6(c^{\dagger}_{n}c_{n}c^{\dagger}_{{m}}c_{{m}}+c^{\dagger}_{\bar{n}}c_{\bar{n}}c^{\dagger}_{\bar{m}}c_{\bar{m}})+\tilde{g}_6' c^{\dagger}_{n}c_{n}c^{\dagger}_{\bar{m}}c_{\bar{m}}
\Big)\Big],\;\;
\label{eq:Hg_i}
\eea
where $c^{\dagger}_{n}$ creates spin-up electrons in patch $n=1,2,3$ at valley $K$, and $c^{\dagger}_{\bar{n}}$ creates spin-down electrons in patch $\bar{n}$ at valley $K'$. 
The strengths of the intra-valley (inter-valley) density-density interactions $\tilde{g}_6$ ($\tilde{g}_2$ and $\tilde{g}_6'$) as well as the inter-valley scattering  $\tilde{g}_3$ in Eq. \ref{eq:Hg_i} are related to the Coulomb potentials in Eq. \ref{eq:Coulomb} as  
\bea
&&
\tilde{g}_2 =\tilde{g}_6'= {V}_{ll}(0)- 
\Lambda_1(D)({V}_{ll}(0)-{V}_{l\bar{l}}(0)),\quad
\nn\\
&&
\tilde{g}_3 =
V_{ll}(Q)
\Lambda_2(D)
+
V_{l\bar{l}}(Q)
\Lambda_3(D),
\nn\\
&&
\tilde{g}_6=\tilde{g}_2-\tilde{g}_3, 
\label{eq:g_and_V}
\eea
where $Q\equiv Q_{nm}=|\textbf{P}_n-\textbf{P}_m|$ for all patches $n$ and $m$,  
and 
\bea
&&
\Lambda_1(D)=|\Lambda_{\uparrow,t}(\tbf{P}_1,\tbf{P}_1) \Lambda_{\uparrow,b}(\tbf{P}_1,\tbf{P}_1) |,
\nn\\
&&
\Lambda_2(D)=|\Lambda_{\uparrow,t}(\tbf{P}_1,\tbf{P}_2)|^2+
|\Lambda_{\uparrow,b}(\tbf{P}_1,\tbf{P}_2)|^2,
\nn\\
&&
\Lambda_3(D)=2|\Lambda_{\uparrow,t}(\tbf{P}_1,\tbf{P}_2)\Lambda_{\uparrow,b}(\tbf{P}_1,\tbf{P}_2)|.
\label{eq:g_D_dependence}
\eea
Using Eqs. \ref{eq:g_and_V} and \ref{eq:g_D_dependence}, we numerically calculate the $D$-dependence of the bare interactions $\tilde{g}_i$ in Fig. \ref{fig:RG_plot}a. 
In particular, since  $\Lambda_{1 (2,3)}$ are decreasing (increasing) functions of $D$ and ${V}_{ll}(0)-{V}_{l\bar{l}}(0)>0$, we find that $\tilde{g}_6'$ and $\tilde{g}_2$ decrease with $D$ while $\tilde{g}_3$ and $\tilde{g}_6$ increase with $D$ (see Fig. \ref{fig:RG_plot}a). 

\textit{Renormalization group analysis ---} 
The inter-patch interactions in Eq. \ref{eq:Hg_i} can lead to a wide variety of FS instabilities even in the weak-to-intermediate-coupling regime since the DOS diverges at VH filling. 
To identify the dominant FS instability at different displacement field strengths $D$, we perform a renormalization group (RG) analysis, which  unbiasly considers the full set of instabilities and identifies the one that is driven by the most relevant interaction in the low-energy limit \cite{RevModPhys.66.129, PhysRevLett.81.3195, HUR20091452, PhysRevB.78.134512, Nandkishore2012}. This approach was widely applied to various few-layer Van der Waals systems in the weak-to-intermediate coupling regime \cite{
PhysRevX.8.041041,
PhysRevB.100.085136, PhysRevB.102.085103, 
PhysRevB.102.125141,
PhysRevB.102.245122, PhysRevB.106.155115, PhysRevB.100.085136, PhysRevB.111.115144}. 
In particular, the framework for general 2D hexagonal systems \textit{without spin degeneracy} was proposed in Ref. \cite{PhysRevB.104.195134}, which consists of three steps. We will briefly review this established method in the following, where details are included in the SM section II for completeness \cite{SM_tTMD}.  

In Step 1, we examine the evolution of the inter-patch coupling constants $g_i(y)$ as a function of an inverse energy scale $y\equiv \nu_0\text{ln}^2(\Lambda/E)/2$~\footnote{The prefactor $\nu_0$ depends on the specific dispersion obtained from Eq. \ref{eq:Moire_H}. {For the RG calculation, we take the phenomenological value $\nu_0=\frac{1}{2\pi W}$ for the bandwidth $W$.}}, which flows from a higher energy scale $\Lambda$ at $y=0$,  corresponding to the patch size $k_{\Lambda}$, towards the low-energy limit at $y\rightarrow\infty$. Such evolution is governed by one-loop RG equations 
with a general form of 
\begin{align}
    \frac{dg_i}{dy}=\sum_{lm}
    d_{lm}g_{l}g_{m}, ~~~ i,l,m=2,3,6,6'. 
    \label{eq:dgdy}
\end{align}   
Here, $g_i\equiv\nu_0\tilde{g}_i$, and the factor $d_{lm}=\frac{1}{\nu_0}\frac{d}{dy}\Pi(\textbf{q})$ captures the evolution of the non-interacting  susceptibilities $\Pi(\textbf{q})$ at momenta $\textbf{q}=0, \textbf{Q}_{nm}$,   which physically correspond to the density of states or the particle-hole and particle-particle nesting degrees between patches $n$ and $m$, respectively.
This set of RG equations is capable of capturing weak-to-intermediate coupling physics in systems with no spin-degeneracy and no perfectly nested FS by including the subleading ln-divergent terms on top of the leading  ln$^2$ terms \cite{PhysRevB.104.195134}. 

 
In Step 2, we identify the  interaction $\Gamma_j$ that supports each FS instability $j$. We consider all symmetry-allowed candidate order parameters $\Delta_j$, including zero-momentum superconductivity $\Delta_j(\textbf{p}) = \langle c^{\dagger}_{\uparrow}(\textbf{p})c^{\dagger}_{\downarrow}(-\textbf{p})\rangle$, pair-density waves $\langle c^{\dagger}_{\uparrow}(\textbf{p})c^{\dagger}_{\uparrow}(-\textbf{p}+\textbf{Q}_{nm})\rangle$, $\langle c^{\dagger}_{\uparrow}(\textbf{p})c^{\dagger}_{\downarrow}(-\textbf{p}+\textbf{Q}_{n\bar{m}})\rangle$, density waves $\langle c^{\dagger}_{s}(\textbf{p})\sigma^{a}_{ss'}c_{s'}(\textbf{p}+\textbf{q})\rangle$, as well as  uniform spin and charge orders $\langle c^{\dagger}_{s}(\textbf{p})\sigma^{a}_{ss'}c_{s'}(\textbf{p})\rangle$, where $\sigma^{a}$ with $a=0,1,2,3$ are the spin Pauli matrices. 
The corresponding interaction $\Gamma_j(y)=\sum_{ji}\eta_{ji}g_i(y)$ that drives each instability $j$ is given by a certain linear combination of the inter-patch interactions $g_i$, which can be systematically identified diagrammatically (see SM Section II \cite{SM_tTMD}). 

In Step 3, we identify the most relevant FS instability $j$ by identifying the strongest tendency $\beta_j\equiv d_j(y)\Gamma_j(y)$  as the inverse energy $y$ flows towards a critical low-energy limit $y_c$, quantified by the product of the driving interaction $\Gamma_j$ and the factor $d_j(y)=\frac{1}{\nu_0}\frac{d}{dy}\Pi(\textbf{q})$ that captures the ratio between the non-interacting susceptibilities of instability $j$ and superconductivity. This tendency determines the susceptibility $\chi_{j}\sim(y_c-y)^{2\beta_j+1}$ of instability $j$, where the dominant instability has the most negative exponent $2\beta_j+1<0$. 

\textit{Displacement-field-driven phase transition --- }
We now perform the RG analysis on the patch model $H_{\text{patch}}$ to show how a displacement field $D$ can drive a superconductivity-magnetism transition in generic 2D hexagonal systems at Van Hove filling. 
Within the framework of patch RG analysis, the dependence of $D$ explicitly enters the initial conditions {$\tilde{g}_i\equiv g_i(y=0)/\nu_0$} of the RG differential equations in Step 1. As shown in Fig. \ref{fig:RG_plot}a and b, the initial conditions $\tilde{g}_{2,3,6,6'}$ exhibit different field-dependence through the Bloch-function overlaps at $y=0$ (see Eqs. \ref{eq:g_and_V},  \ref{eq:g_D_dependence}) so that 
the signs and relevance of $g_i(y)$ in the low-energy limit  $y\rightarrow\infty$, and thus the instability tendency $\beta_j$, can qualitatively alter as $D$ increases. 
A change in the dominant instability $j$ can thus occur when $D$ reaches some critical value $D_c$, hence a displacement-field driven phase transition.

By numerically calculating the tendencies $\beta_j$ for all  instabilities $j$ considered in Step 2, we find that the two leading instabilities are the 
$d/p$-wave superconductivity ($d/p$-SC) and the valley ferromagnetism (vFM) when the parameters for tWSe$_2$ are used. 
Specifically, their corresponding tendencies $\beta_j$ are given by \cite{PhysRevB.104.195134}
\bea
&&
\beta_{d/p\text{-SC}} = G_2-G_3,
\nn\\
&&
\beta_{\text{vFM}} = -d(y)(G_2-2G_6+2G_6'), 
\label{eq:beta}
\eea 
where $d(y)=\frac{d}{dy}\Pi_{\text{ph}}(0)$. 
In Fig. \ref{fig:RG_plot}b, we show the competition between these two leading instabilities quantified by $2\beta_j+1$, where 
we find that the $d/p$-SC phase dominates at a weaker field $D<D_c$, whereas the vFM phase is favored at a stronger field $D>D_c$ through a Stoner-like transition.    

\begin{figure*}[t]
\centering
{\label{fig:plot}
\includegraphics[width=0.91\textwidth]{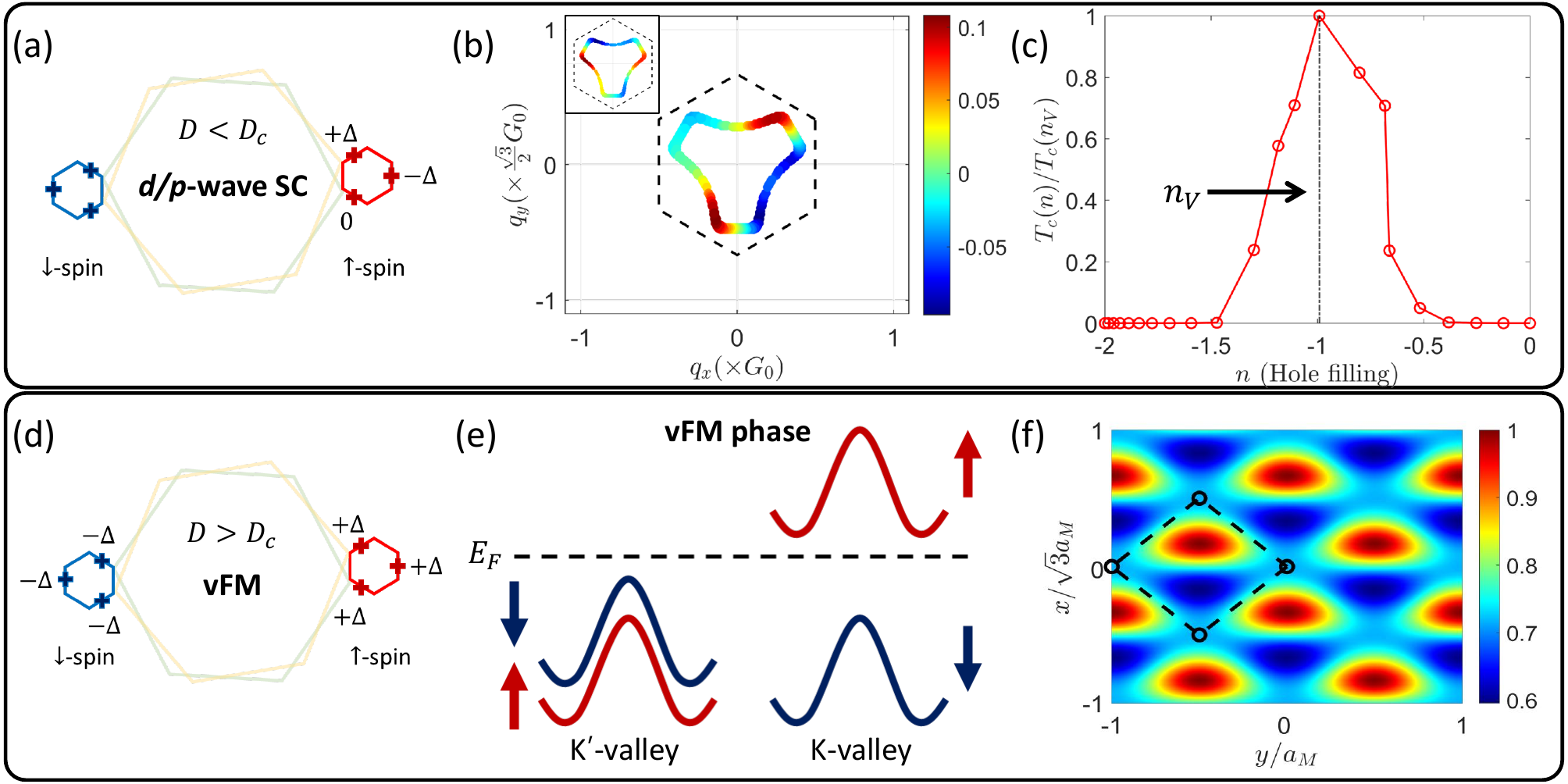}}
\caption{
(a) Graphical representation of the mixture of $p$- and $d$-wave superconducting pairings.
(b) The gap function, $\Delta(\tbf{k})$ obtained by solving the linearized gap equation, Eq. \ref{eq:gap_eq} on the FS. It belongs to $E$-representation of $C_{3v}$-symmetry group and the other solution (inset) is obtained by applying 3-fold rotational transformation.
(c) The behavior of $T_c(n)/T_c(n_V)$ is plotted by solving Eq. \ref{eq:gap_eq} where $n_{V}\approx 1$ is the VH-filling at $D=30\text{meV}$ (along the dashed blue line in Fig. \ref{fig:schematic}a). 
For the numerical plot, we adapt the bare Hubbard interaction, $U=109\text{meV}$ \cite{Guo2025}.
(d) Graphical representation of the ferromagnetic order parameter.
(e) Spin density, $m(\tbf{r})=\la \psi^\dagger_{\uparrow}(\tbf{r})\psi_\uparrow(\tbf{r})-\psi^\dagger_{\downarrow}(\tbf{r})\psi_{\downarrow}(\tbf{r}) \ra$ in the real space, $\tbf{r}=(x,y)$ at the VH-filling in the vFM phase with $\Delta_{\text{vFM}}=50\text{meV}$ and $D=40\text{meV}$. The moire unit cell is marked by dotted line and the intensity peak is normalized to be $1$.
(f) Reconstructed Fermi surface of vFM phase for up-spin (red) and down-spin (blue) electrons in $K$ and $K'$-valleys. The Fermi level $E_F$ in the normal state is marked.
}
\label{fig:plot}
\end{figure*} 

We now  discuss  how the $D$-dependence in the initial conditions at a higher-energy $y=0$ in Fig. \ref{fig:RG_plot}a drives the transition from $d/p$-SC to vFM in the low-energy limit in Fig. \ref{fig:RG_plot}b. 
Under a weak field $D<D_c$ where SC dominates, the sign and relevance of the inter-patch interactions $g_i$ are shown in the RG flows in Fig. \ref{fig:RG_plot}c.  
The density-density interaction $g_2$ between  opposite patches $\textbf{P}_n$ and $\textbf{P}_{\bar{n}}$ is the key interaction that drives the Kohn-Luttinger-like SC. 
Although starting out as a repulsion at the higher energy scale $y=0$, $g_2$ becomes attractive due to the particle-hole fluctuations from other interactions in the low-energy limit. 
Moreover, we find that the unconventional $d/p$-wave pairing dominates over the conventional pairing driven by $G_2+2G_3$ (See SM Section II \cite{SM_tTMD}) due to the repulsively relevant inter-patch scattering $g_3$, which favors momentum-dependent order parameters. 

As the displacement field grows stronger to  $D>D_c$, 
the typical RG flows of $g_i(y)$ qualitatively change from Fig. \ref{fig:RG_plot}c to Fig. \ref{fig:RG_plot}d due to the seemingly mild changes in the initial condition, which now favor vFM over superconductivity. 
There are two key changes in the RG flows that lead to the phase transition: (1) An inter-valley density density interaction $g_2$ flips from a relevant attraction that drives SC into a repulsion that drives FM instead. Second,  the intra-valley density-density interaction $g_6$ flips from a relevant repulsion that hurts vFM into an attraction that supports vFM (see Eq. \ref{eq:beta}).  
These two changes collectively drive a Stoner-like transition when the displacement field $D$ becomes  stronger.  


\textit{Topological  superconductivity at $D<D_c$ ---} 
Our RG results at a weaker field finds that the superconductivity favors two degenerate  nodal $d/p$-wave pairing gaps (see Fig. \ref{fig:plot}a), which form a two-dimensional irreducible representation (irrep) in the relevant point group $C_{3v}$. Due to energetic considerations \cite{PhysRevB.110.035143,Nandkishore2012}, these two degenerate gaps are known to prefer a chiral linear combination so that the FS can be fully gapped (See SM Section II \cite{SM_tTMD}). We therefore expect a chiral $d/p$-wave SC that is topological with Chern number 2.  

Although our RG result was obtained at VH filling, the finding of chiral $d/p$-wave pairing gap is robust even away from the VH filling in spin-orbit coupled hexagonal systems, assuming that the screened Coulomb interaction is stronger than electron-phonon coupling. This is because such systems typically has a point group of $C_{3v}$, where the inversion symmetry is broken by the spin-orbit coupling so that the even- and odd-parity gaps can mix. There are thus only two possible fully gapped superconducting states described by the irreps of $C_{3v}$: the fully gapped $s/f$-wave gap in the trivial irrep A, and our case of chiral $d/p$-wave gap in the two-dimensional irrep E. Due to the $s$-wave component in the former possibility, screened Coulomb interaction generally favors the chiral $d/p$-wave gap. 

To further demonstrate the stability of the chiral $d/p$-wave SC away from the Van Hove filling $n_{V}(D)$, in Fig. \ref{fig:plot}b,c we calculate the gap function $\Delta(\tbf{p})$ on the full FS and the critical temperature $T_c(n)/T_c(n_V)$ along the black dotted line in Fig. \ref{fig:schematic}a by 
solving the linearized gap equation self-consistently 
\bea
\int_{\tbf{p}'\in\text{FS}}
\frac{d\hat{\tbf{p}}'}{|\tbf{v}_{\tbf{p}'}|}\Gamma(\tbf{p},\tbf{p}') \Delta(\tbf{p}')  = \lambda \Delta(\tbf{p}), 
\label{eq:gap_eq}
\eea 
where $\tbf{v}_\tbf{p}$ is the Fermi velocity, the interaction vertex  $\Gamma(\textbf{p},\textbf{p}')$ is set to be the dressed Hubbard interaction up to one-loop correction \cite{Hsu2017}, and the effective pairing interaction strength $\lambda$ is obtained by solving for the eigenvalues of $\Gamma(\tbf{p},\tbf{p}')$, which determines the critical temperature $T_c$. 
In Fig. \ref{fig:plot}b, we find that the energetically favored gap functions remain in the $E$ irrep when moving away from the VH filling, where we show the two degenerate $d/p$-wave gap functions on the FS {at the representative point labeled in Fig. \ref{fig:schematic}a}. 
Moreover, we show that this $d/p$-wave pairing gap remains stable away from the VH filling, as {signaled by the smallest effective interaction $\lambda_{\text{min}}<0$ being negative (see Fig. \ref{fig:plot}c)}.
This effective pairing interaction determines the critical temperature as  
$T_c \propto e^{-1/|\lambda_{\text{min}}|}$, which peaks at the VH filling $n_V$ as expected from the diverging DOS.

\textit{Valley ferromagnetism at $D>D_c$---} 
The vFM phase we find at a stronger field $D>D_c$ is characterized by the order parameter (see Fig. \ref{fig:plot}d) 
\bea
\delta H_{\text{vFM}} = \Delta_{\text{vFM}}
\Big(
\sum_{\tbf{p}\in K}c_{\uparrow}^\dagger(\tbf{p})c_{\uparrow}(\tbf{p})
-\sum_{\tbf{p}\in K'}
c_{\downarrow}^\dagger(\tbf{p})c_{\downarrow}(\tbf{p})
\Big).  \;\; 
\label{eq:H_fFM}
\eea
Due to the strong Ising SOC, this ferromagnetic order parameter pushes the topmost bands at valley $K$ and $K'$ up and down, respectively (see Fig. \ref{fig:plot}e), resulting in a vFM state. 
Intuitively, this order parameter acts as a time-reversal-breaking pseudo-magnetic field that enhances (suppresses) the Ising SOC strength at valley $K$ ($K'$), leading to a \textit{finite magnetization that is spatially non-uniform} (see Fig. \ref{fig:plot}f).

\begin{figure}[t!]
{\label{fig:RG}
\includegraphics[width=0.5\textwidth]{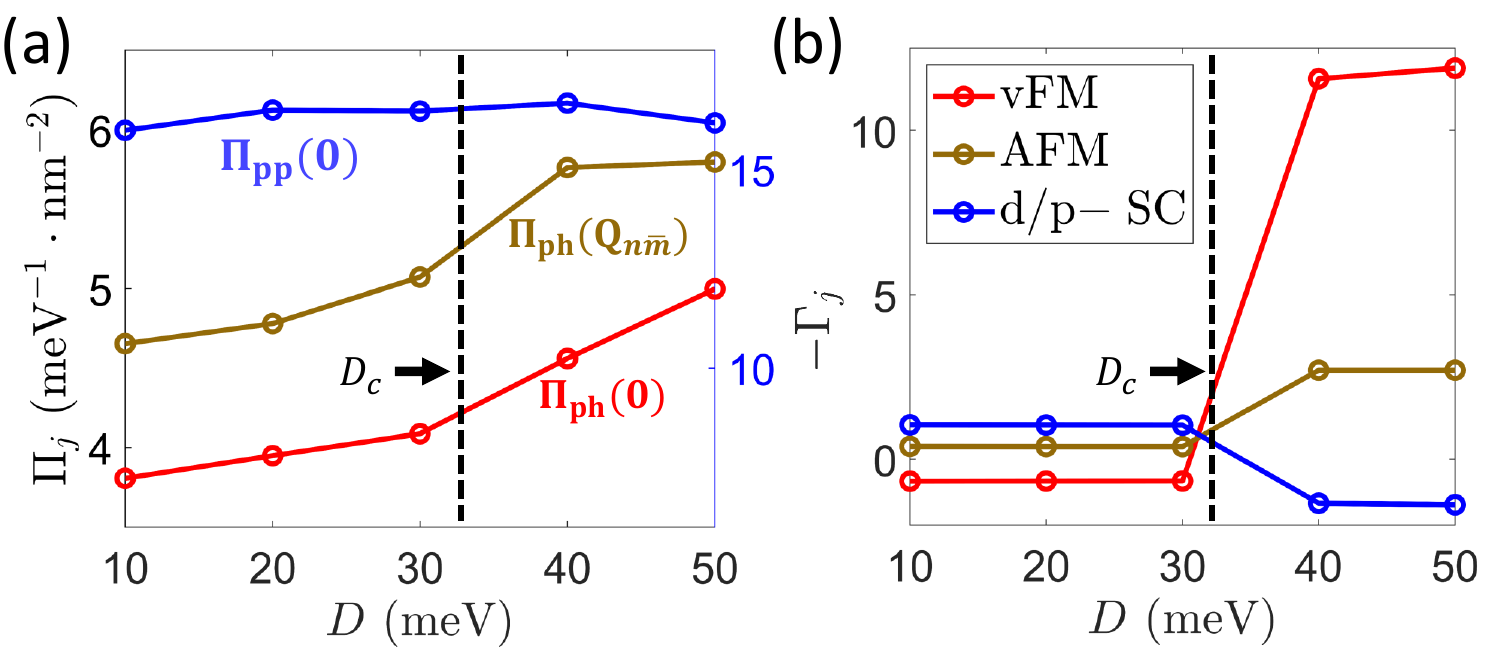}}
\caption{
(a) Non-interacting susceptibility (only $\Pi_{\text{pp}}(0)$-data is assigned to the right y-axis) and (b) driving interaction for instability $j$ as a function of the displacement field $D$ at the critical inverse energy scale $y_c$. At $D$ larger than the transition field strength $D_c$, the non-interacting susceptibilities of the AFM state $\Pi_{\text{ph}}(\textbf{Q}_{n\bar{m}})$ in (a) is slightly larger than that of the vFM state $\Pi_{\text{ph}}(0)$ due to the stronger nesting degree between the two valleys. Nonetheless, in (b), the driving interactions of the vFM state is larger than the AFM state. The AFM order parameter is given by $\Delta_{\text{AFM}}=\la c^\dagger_{\uparrow}(\tbf{p})c_{\downarrow}(\tbf{p}+\tbf{Q}_{n\bar{m}}) + c^\dagger_{\uparrow}(\tbf{p})c_{\downarrow}(\tbf{p}+\tbf{Q}_{m\bar{n}}) \ra$.  
}
\label{fig:sus_plot}
\end{figure}

We now comment on the difference and competition between the vFM state we find and  antiferromagnetic (AFM) states that were often proposed in prior works for twisted bilayer TMDs \cite{tuo2024theorytopologicalsuperconductivityantiferromagnetic,fischer2024theoryintervalleycoherentafmorder}. 
Although both vFM and AFM order parameters are spatially non-uniform,  
vFM has a finite magnetization while AFM has zero magnetization. 
In the vFM state, the magnetization originates from the imbalance between spin densities between the spin-up density at valley $K$ and the spin-down density at valley $K'$, and thus spatially modulates  at a wavelength $\propto  \pm 1/(2K)$.  
This vFM instability can naturally occur when a spin ferromagnetic order develops in an Ising spin-orbit-coupled metal with multiple valleys. 
The competition between the vFM and AFM instabilities can be quantified by the tendency  $\beta_j\sim\Pi_j\Gamma_j$, where $\Pi_j$ and $\Gamma_j$ are  the corresponding non-interacting susceptibility and the driving interaction of instability $j$ at the critical inverse energy scale $y_c$, respectively \cite{RG_DTBG}. As shown in Fig. \ref{fig:sus_plot}, although the non-interacting susceptibility of the AFM state $\Pi_j=\Pi_{\text{ph}}(\tbf{Q}_{n\bar{m}})$ is slightly larger than that of the vFM state $\Pi_{\text{ph}}(0)$ at a larger displacement field $D>D_c$, the driving interaction of the vFM state $\Gamma_{\text{vFM}}$ is much larger than the AFM interaction $\Gamma_{\text{AFM}}$ so that the system tilts towards the vFM state in this close competition.  

\textit{Experimental detections and material systems ---}
The proposed vFM state can be experimentally distinguished from an AFM state by detecting its finite magnetization, e.g. using  magnetic dichroism, as was done in $2^{\circ}$ to $3^{\circ}$ tWSe$_2$ \cite{Knüppel2025}. 
Furthermore, the valley-dependent band structures in Fig. \ref{fig:plot}e could be measured by angle-resolved photoemission spectroscopy (ARPES), whereas the non-uniform {spin density} in Fig. \ref{fig:plot}f could be detected by spin-dependent local probes, such as spin-dependent scanning tunneling microscope or scanning  superconducting quantum interference device (SQUID). Finally, with transport measurements, 
ferromagnetism and AFM  can be distinguished from their Curie–Weiss behaviors in temperature dependence  \cite{Knüppel2025, Tang2020}. 
The resistivity of the vFM state depends on the filling due to the relation between the reconstructed low-energy band structure and chemical potential, which could reach an insulating state when the vFM gap is comparable to the bandwidth. 
As shown in Fig. \ref{fig:plot}e, the vFM state becomes strictly insulating at filling $n=1\sim n_V(D=40\text{meV})$,  and the longitudinal resistivity may slowly decrease as moving away from $n=1$.


The nature of the superconducting state can be experimentally examined by several ways. First of all, the hypothesis that the superconductivity is driven by multiple VHS on the FS can be validated by an optical absorption measurement \textit{in the superconducting state}, signaled by a pronounced peak at frequency $\omega\sim 2\Delta$ exceeding the disorder background \cite{PhysRevB.111.054514}. The anisotropy in the current-induced optical conductivity Re$\sigma_{ii}(\omega)$ between $i=x$ and $y$ can further determine whether these VHS are conventional and higher-order, with logarithmic and power-law diverging DOS, respectively  \cite{PhysRevB.111.054514}.  For the pairing symmetry, the conventional $s/f$-wave and unconventional $d/p$-wave order parameters can be distinguished with phase-sensitive measurements \cite{Hsu_monoTMD}.  

Finally, we comment on the material platforms that could realize the proposed  phase transition as well as the valley ferromagnetism and chiral $d/p$-wave superconductivity on the two sides of the transition. One plausible system that exhibits a similar displacement-field-driven superconducting to magnetic phase transition is the $5^\circ$-tWSe$_2$ \cite{Guo2025}, from which we adapt the material parameters for our numerical calculations.  
We propose that the observed SC to magnetism transition in Ref. \cite{Guo2025} can be a chiral $d/p$-wave SC-to-vFM transition mediated by inter-VH interactions, if the FS away from VH points does not exhibit perfect nesting, as we find using the continuum model $H_0$ \cite{PhysRevLett.122.086402}. 
While previous theoretical works have proposed magnetic states with large momentum transfer, such as 120-degree antiferromagnetism \cite{tuo2024theorytopologicalsuperconductivityantiferromagnetic} and inter-valley coherent states \cite{fischer2024theoryintervalleycoherentafmorder}, definitive experimental evidence for the magnetic order parameter remains elusive. We thus urge near-future detections to examine our proposal.

Besides the displacement field driven transitions, we expect that a similar Stoner-like transition into the vFM phase could also be driven by other experimental knobs that can effectively tune the interaction strengths among VH points near the VH filling, such as the twist angle. Recent experimental studies in different smaller-angle tWSe$_2$ systems have reported ferromagnetism \cite{Knüppel2025,gao2025probingquantumanomaloushall}, offering a promising platform to test our proposed vFM state via the suggested measurements.  Finally, our results highlight that such a Stoner-like transition into a vFM state and a nearby chiral $d/p$-wave SC are generic  weak-to-intermediate coupling phenomenon that can occur in other twisted or untwisted spin-orbit-coupled few-layer systems near the VH filling away from perfect nesting.

\textit{Acknowledgment---} The authors are grateful for useful discussions with Fengcheng Wu, Ming Xie, and Daniel Kaplan. H.-J. Y. and Y.-T. H. acknowledge the support from National Science Foundation Grant
No. DMR-2238748. H.-J.Y. also acknowledges the support from the Society of Science Fellows Postdoctoral Program in the College of Science. 
This work was performed in part at Aspen Center for Physics, which is supported by National Science Foundation grant PHY-2210452.  This research was also supported in part by grant NSF PHY-2309135 to the Kavli Institute for Theoretical Physics (KITP). 

\bibliography{ref}

\begin{thebibliography}{66}%
\makeatletter
\providecommand \@ifxundefined [1]{%
 \@ifx{#1\undefined}
}%
\providecommand \@ifnum [1]{%
 \ifnum #1\expandafter \@firstoftwo
 \else \expandafter \@secondoftwo
 \fi
}%
\providecommand \@ifx [1]{%
 \ifx #1\expandafter \@firstoftwo
 \else \expandafter \@secondoftwo
 \fi
}%
\providecommand \natexlab [1]{#1}%
\providecommand \enquote  [1]{``#1''}%
\providecommand \bibnamefont  [1]{#1}%
\providecommand \bibfnamefont [1]{#1}%
\providecommand \citenamefont [1]{#1}%
\providecommand \href@noop [0]{\@secondoftwo}%
\providecommand \href [0]{\begingroup \@sanitize@url \@href}%
\providecommand \@href[1]{\@@startlink{#1}\@@href}%
\providecommand \@@href[1]{\endgroup#1\@@endlink}%
\providecommand \@sanitize@url [0]{\catcode `\\12\catcode `\$12\catcode `\&12\catcode `\#12\catcode `\^12\catcode `\_12\catcode `\%12\relax}%
\providecommand \@@startlink[1]{}%
\providecommand \@@endlink[0]{}%
\providecommand \url  [0]{\begingroup\@sanitize@url \@url }%
\providecommand \@url [1]{\endgroup\@href {#1}{\urlprefix }}%
\providecommand \urlprefix  [0]{URL }%
\providecommand \Eprint [0]{\href }%
\providecommand \doibase [0]{https://doi.org/}%
\providecommand \selectlanguage [0]{\@gobble}%
\providecommand \bibinfo  [0]{\@secondoftwo}%
\providecommand \bibfield  [0]{\@secondoftwo}%
\providecommand \translation [1]{[#1]}%
\providecommand \BibitemOpen [0]{}%
\providecommand \bibitemStop [0]{}%
\providecommand \bibitemNoStop [0]{.\EOS\space}%
\providecommand \EOS [0]{\spacefactor3000\relax}%
\providecommand \BibitemShut  [1]{\csname bibitem#1\endcsname}%
\let\auto@bib@innerbib\@empty
\bibitem [{\citenamefont {Kim}\ \emph {et~al.}(2017)\citenamefont {Kim}, \citenamefont {DaSilva}, \citenamefont {Huang}, \citenamefont {Fallahazad}, \citenamefont {Larentis}, \citenamefont {Taniguchi}, \citenamefont {Watanabe}, \citenamefont {LeRoy}, \citenamefont {MacDonald},\ and\ \citenamefont {Tutuc}}]{doi:10.1073/pnas.1620140114}%
  \BibitemOpen
  \bibfield  {author} {\bibinfo {author} {\bibfnamefont {K.}~\bibnamefont {Kim}}, \bibinfo {author} {\bibfnamefont {A.}~\bibnamefont {DaSilva}}, \bibinfo {author} {\bibfnamefont {S.}~\bibnamefont {Huang}}, \bibinfo {author} {\bibfnamefont {B.}~\bibnamefont {Fallahazad}}, \bibinfo {author} {\bibfnamefont {S.}~\bibnamefont {Larentis}}, \bibinfo {author} {\bibfnamefont {T.}~\bibnamefont {Taniguchi}}, \bibinfo {author} {\bibfnamefont {K.}~\bibnamefont {Watanabe}}, \bibinfo {author} {\bibfnamefont {B.~J.}\ \bibnamefont {LeRoy}}, \bibinfo {author} {\bibfnamefont {A.~H.}\ \bibnamefont {MacDonald}},\ and\ \bibinfo {author} {\bibfnamefont {E.}~\bibnamefont {Tutuc}},\ }\bibfield  {title} {\bibinfo {title} {Tunable moiré bands and strong correlations in small-twist-angle bilayer graphene},\ }\href {https://doi.org/10.1073/pnas.1620140114} {\bibfield  {journal} {\bibinfo  {journal} {Proceedings of the National Academy of Sciences}\ }\textbf {\bibinfo {volume} {114}},\ \bibinfo {pages} {3364} (\bibinfo {year} {2017})},\
  \Eprint {https://arxiv.org/abs/https://www.pnas.org/doi/pdf/10.1073/pnas.1620140114} {https://www.pnas.org/doi/pdf/10.1073/pnas.1620140114} \BibitemShut {NoStop}%
\bibitem [{\citenamefont {Carr}\ \emph {et~al.}(2017)\citenamefont {Carr}, \citenamefont {Massatt}, \citenamefont {Fang}, \citenamefont {Cazeaux}, \citenamefont {Luskin},\ and\ \citenamefont {Kaxiras}}]{PhysRevB.95.075420}%
  \BibitemOpen
  \bibfield  {author} {\bibinfo {author} {\bibfnamefont {S.}~\bibnamefont {Carr}}, \bibinfo {author} {\bibfnamefont {D.}~\bibnamefont {Massatt}}, \bibinfo {author} {\bibfnamefont {S.}~\bibnamefont {Fang}}, \bibinfo {author} {\bibfnamefont {P.}~\bibnamefont {Cazeaux}}, \bibinfo {author} {\bibfnamefont {M.}~\bibnamefont {Luskin}},\ and\ \bibinfo {author} {\bibfnamefont {E.}~\bibnamefont {Kaxiras}},\ }\bibfield  {title} {\bibinfo {title} {Twistronics: Manipulating the electronic properties of two-dimensional layered structures through their twist angle},\ }\href {https://doi.org/10.1103/PhysRevB.95.075420} {\bibfield  {journal} {\bibinfo  {journal} {Phys. Rev. B}\ }\textbf {\bibinfo {volume} {95}},\ \bibinfo {pages} {075420} (\bibinfo {year} {2017})}\BibitemShut {NoStop}%
\bibitem [{\citenamefont {Wu}\ \emph {et~al.}(2018)\citenamefont {Wu}, \citenamefont {Lovorn}, \citenamefont {Tutuc},\ and\ \citenamefont {MacDonald}}]{PhysRevLett.121.026402}%
  \BibitemOpen
  \bibfield  {author} {\bibinfo {author} {\bibfnamefont {F.}~\bibnamefont {Wu}}, \bibinfo {author} {\bibfnamefont {T.}~\bibnamefont {Lovorn}}, \bibinfo {author} {\bibfnamefont {E.}~\bibnamefont {Tutuc}},\ and\ \bibinfo {author} {\bibfnamefont {A.~H.}\ \bibnamefont {MacDonald}},\ }\bibfield  {title} {\bibinfo {title} {Hubbard model physics in transition metal dichalcogenide moir\'e bands},\ }\href {https://doi.org/10.1103/PhysRevLett.121.026402} {\bibfield  {journal} {\bibinfo  {journal} {Phys. Rev. Lett.}\ }\textbf {\bibinfo {volume} {121}},\ \bibinfo {pages} {026402} (\bibinfo {year} {2018})}\BibitemShut {NoStop}%
\bibitem [{\citenamefont {Chittari}\ \emph {et~al.}(2019)\citenamefont {Chittari}, \citenamefont {Chen}, \citenamefont {Zhang}, \citenamefont {Wang},\ and\ \citenamefont {Jung}}]{PhysRevLett.122.016401}%
  \BibitemOpen
  \bibfield  {author} {\bibinfo {author} {\bibfnamefont {B.~L.}\ \bibnamefont {Chittari}}, \bibinfo {author} {\bibfnamefont {G.}~\bibnamefont {Chen}}, \bibinfo {author} {\bibfnamefont {Y.}~\bibnamefont {Zhang}}, \bibinfo {author} {\bibfnamefont {F.}~\bibnamefont {Wang}},\ and\ \bibinfo {author} {\bibfnamefont {J.}~\bibnamefont {Jung}},\ }\bibfield  {title} {\bibinfo {title} {Gate-tunable topological flat bands in trilayer graphene boron-nitride moir\'e superlattices},\ }\href {https://doi.org/10.1103/PhysRevLett.122.016401} {\bibfield  {journal} {\bibinfo  {journal} {Phys. Rev. Lett.}\ }\textbf {\bibinfo {volume} {122}},\ \bibinfo {pages} {016401} (\bibinfo {year} {2019})}\BibitemShut {NoStop}%
\bibitem [{\citenamefont {Chen}\ \emph {et~al.}(2019{\natexlab{a}})\citenamefont {Chen}, \citenamefont {Jiang}, \citenamefont {Wu}, \citenamefont {Lyu}, \citenamefont {Li}, \citenamefont {Chittari}, \citenamefont {Watanabe}, \citenamefont {Taniguchi}, \citenamefont {Shi}, \citenamefont {Jung}, \citenamefont {Zhang},\ and\ \citenamefont {Wang}}]{Chen2019}%
  \BibitemOpen
  \bibfield  {author} {\bibinfo {author} {\bibfnamefont {G.}~\bibnamefont {Chen}}, \bibinfo {author} {\bibfnamefont {L.}~\bibnamefont {Jiang}}, \bibinfo {author} {\bibfnamefont {S.}~\bibnamefont {Wu}}, \bibinfo {author} {\bibfnamefont {B.}~\bibnamefont {Lyu}}, \bibinfo {author} {\bibfnamefont {H.}~\bibnamefont {Li}}, \bibinfo {author} {\bibfnamefont {B.~L.}\ \bibnamefont {Chittari}}, \bibinfo {author} {\bibfnamefont {K.}~\bibnamefont {Watanabe}}, \bibinfo {author} {\bibfnamefont {T.}~\bibnamefont {Taniguchi}}, \bibinfo {author} {\bibfnamefont {Z.}~\bibnamefont {Shi}}, \bibinfo {author} {\bibfnamefont {J.}~\bibnamefont {Jung}}, \bibinfo {author} {\bibfnamefont {Y.}~\bibnamefont {Zhang}},\ and\ \bibinfo {author} {\bibfnamefont {F.}~\bibnamefont {Wang}},\ }\bibfield  {title} {\bibinfo {title} {Evidence of a gate-tunable mott insulator in a trilayer graphene moir{\'e} superlattice},\ }\href {https://doi.org/10.1038/s41567-018-0387-2} {\bibfield  {journal} {\bibinfo  {journal} {Nature Physics}\ }\textbf {\bibinfo
  {volume} {15}},\ \bibinfo {pages} {237} (\bibinfo {year} {2019}{\natexlab{a}})}\BibitemShut {NoStop}%
\bibitem [{\citenamefont {Lu}\ \emph {et~al.}(2019)\citenamefont {Lu}, \citenamefont {Stepanov}, \citenamefont {Yang}, \citenamefont {Xie}, \citenamefont {Aamir}, \citenamefont {Das}, \citenamefont {Urgell}, \citenamefont {Watanabe}, \citenamefont {Taniguchi}, \citenamefont {Zhang}, \citenamefont {Bachtold}, \citenamefont {MacDonald},\ and\ \citenamefont {Efetov}}]{Lu2019}%
  \BibitemOpen
  \bibfield  {author} {\bibinfo {author} {\bibfnamefont {X.}~\bibnamefont {Lu}}, \bibinfo {author} {\bibfnamefont {P.}~\bibnamefont {Stepanov}}, \bibinfo {author} {\bibfnamefont {W.}~\bibnamefont {Yang}}, \bibinfo {author} {\bibfnamefont {M.}~\bibnamefont {Xie}}, \bibinfo {author} {\bibfnamefont {M.~A.}\ \bibnamefont {Aamir}}, \bibinfo {author} {\bibfnamefont {I.}~\bibnamefont {Das}}, \bibinfo {author} {\bibfnamefont {C.}~\bibnamefont {Urgell}}, \bibinfo {author} {\bibfnamefont {K.}~\bibnamefont {Watanabe}}, \bibinfo {author} {\bibfnamefont {T.}~\bibnamefont {Taniguchi}}, \bibinfo {author} {\bibfnamefont {G.}~\bibnamefont {Zhang}}, \bibinfo {author} {\bibfnamefont {A.}~\bibnamefont {Bachtold}}, \bibinfo {author} {\bibfnamefont {A.~H.}\ \bibnamefont {MacDonald}},\ and\ \bibinfo {author} {\bibfnamefont {D.~K.}\ \bibnamefont {Efetov}},\ }\bibfield  {title} {\bibinfo {title} {Superconductors, orbital magnets and correlated states in magic-angle bilayer graphene},\ }\href
  {https://doi.org/10.1038/s41586-019-1695-0} {\bibfield  {journal} {\bibinfo  {journal} {Nature}\ }\textbf {\bibinfo {volume} {574}},\ \bibinfo {pages} {653} (\bibinfo {year} {2019})}\BibitemShut {NoStop}%
\bibitem [{\citenamefont {Chen}\ \emph {et~al.}(2019{\natexlab{b}})\citenamefont {Chen}, \citenamefont {Sharpe}, \citenamefont {Gallagher}, \citenamefont {Rosen}, \citenamefont {Fox}, \citenamefont {Jiang}, \citenamefont {Lyu}, \citenamefont {Li}, \citenamefont {Watanabe}, \citenamefont {Taniguchi}, \citenamefont {Jung}, \citenamefont {Shi}, \citenamefont {Goldhaber-Gordon}, \citenamefont {Zhang},\ and\ \citenamefont {Wang}}]{Chen2019_2}%
  \BibitemOpen
  \bibfield  {author} {\bibinfo {author} {\bibfnamefont {G.}~\bibnamefont {Chen}}, \bibinfo {author} {\bibfnamefont {A.~L.}\ \bibnamefont {Sharpe}}, \bibinfo {author} {\bibfnamefont {P.}~\bibnamefont {Gallagher}}, \bibinfo {author} {\bibfnamefont {I.~T.}\ \bibnamefont {Rosen}}, \bibinfo {author} {\bibfnamefont {E.~J.}\ \bibnamefont {Fox}}, \bibinfo {author} {\bibfnamefont {L.}~\bibnamefont {Jiang}}, \bibinfo {author} {\bibfnamefont {B.}~\bibnamefont {Lyu}}, \bibinfo {author} {\bibfnamefont {H.}~\bibnamefont {Li}}, \bibinfo {author} {\bibfnamefont {K.}~\bibnamefont {Watanabe}}, \bibinfo {author} {\bibfnamefont {T.}~\bibnamefont {Taniguchi}}, \bibinfo {author} {\bibfnamefont {J.}~\bibnamefont {Jung}}, \bibinfo {author} {\bibfnamefont {Z.}~\bibnamefont {Shi}}, \bibinfo {author} {\bibfnamefont {D.}~\bibnamefont {Goldhaber-Gordon}}, \bibinfo {author} {\bibfnamefont {Y.}~\bibnamefont {Zhang}},\ and\ \bibinfo {author} {\bibfnamefont {F.}~\bibnamefont {Wang}},\ }\bibfield  {title} {\bibinfo {title} {Signatures of
  tunable superconductivity in a trilayer graphene moir{\'e} superlattice},\ }\href {https://doi.org/10.1038/s41586-019-1393-y} {\bibfield  {journal} {\bibinfo  {journal} {Nature}\ }\textbf {\bibinfo {volume} {572}},\ \bibinfo {pages} {215} (\bibinfo {year} {2019}{\natexlab{b}})}\BibitemShut {NoStop}%
\bibitem [{\citenamefont {Chen}\ \emph {et~al.}(2020)\citenamefont {Chen}, \citenamefont {Sharpe}, \citenamefont {Fox}, \citenamefont {Zhang}, \citenamefont {Wang}, \citenamefont {Jiang}, \citenamefont {Lyu}, \citenamefont {Li}, \citenamefont {Watanabe}, \citenamefont {Taniguchi}, \citenamefont {Shi}, \citenamefont {Senthil}, \citenamefont {Goldhaber-Gordon}, \citenamefont {Zhang},\ and\ \citenamefont {Wang}}]{Chen_2020}%
  \BibitemOpen
  \bibfield  {author} {\bibinfo {author} {\bibfnamefont {G.}~\bibnamefont {Chen}}, \bibinfo {author} {\bibfnamefont {A.~L.}\ \bibnamefont {Sharpe}}, \bibinfo {author} {\bibfnamefont {E.~J.}\ \bibnamefont {Fox}}, \bibinfo {author} {\bibfnamefont {Y.-H.}\ \bibnamefont {Zhang}}, \bibinfo {author} {\bibfnamefont {S.}~\bibnamefont {Wang}}, \bibinfo {author} {\bibfnamefont {L.}~\bibnamefont {Jiang}}, \bibinfo {author} {\bibfnamefont {B.}~\bibnamefont {Lyu}}, \bibinfo {author} {\bibfnamefont {H.}~\bibnamefont {Li}}, \bibinfo {author} {\bibfnamefont {K.}~\bibnamefont {Watanabe}}, \bibinfo {author} {\bibfnamefont {T.}~\bibnamefont {Taniguchi}}, \bibinfo {author} {\bibfnamefont {Z.}~\bibnamefont {Shi}}, \bibinfo {author} {\bibfnamefont {T.}~\bibnamefont {Senthil}}, \bibinfo {author} {\bibfnamefont {D.}~\bibnamefont {Goldhaber-Gordon}}, \bibinfo {author} {\bibfnamefont {Y.}~\bibnamefont {Zhang}},\ and\ \bibinfo {author} {\bibfnamefont {F.}~\bibnamefont {Wang}},\ }\bibfield  {title} {\bibinfo {title} {Tunable correlated
  chern insulator and ferromagnetism in a moiré superlattice},\ }\href {https://doi.org/10.1038/s41586-020-2049-7} {\bibfield  {journal} {\bibinfo  {journal} {Nature}\ }\textbf {\bibinfo {volume} {579}},\ \bibinfo {pages} {56–61} (\bibinfo {year} {2020})}\BibitemShut {NoStop}%
\bibitem [{\citenamefont {Tang}\ \emph {et~al.}(2020)\citenamefont {Tang}, \citenamefont {Li}, \citenamefont {Li}, \citenamefont {Xu}, \citenamefont {Liu}, \citenamefont {Barmak}, \citenamefont {Watanabe}, \citenamefont {Taniguchi}, \citenamefont {MacDonald}, \citenamefont {Shan},\ and\ \citenamefont {Mak}}]{Tang2020}%
  \BibitemOpen
  \bibfield  {author} {\bibinfo {author} {\bibfnamefont {Y.}~\bibnamefont {Tang}}, \bibinfo {author} {\bibfnamefont {L.}~\bibnamefont {Li}}, \bibinfo {author} {\bibfnamefont {T.}~\bibnamefont {Li}}, \bibinfo {author} {\bibfnamefont {Y.}~\bibnamefont {Xu}}, \bibinfo {author} {\bibfnamefont {S.}~\bibnamefont {Liu}}, \bibinfo {author} {\bibfnamefont {K.}~\bibnamefont {Barmak}}, \bibinfo {author} {\bibfnamefont {K.}~\bibnamefont {Watanabe}}, \bibinfo {author} {\bibfnamefont {T.}~\bibnamefont {Taniguchi}}, \bibinfo {author} {\bibfnamefont {A.~H.}\ \bibnamefont {MacDonald}}, \bibinfo {author} {\bibfnamefont {J.}~\bibnamefont {Shan}},\ and\ \bibinfo {author} {\bibfnamefont {K.~F.}\ \bibnamefont {Mak}},\ }\bibfield  {title} {\bibinfo {title} {Simulation of hubbard model physics in wse2/ws2 moir{\'e} superlattices},\ }\href {https://doi.org/10.1038/s41586-020-2085-3} {\bibfield  {journal} {\bibinfo  {journal} {Nature}\ }\textbf {\bibinfo {volume} {579}},\ \bibinfo {pages} {353} (\bibinfo {year} {2020})}\BibitemShut
  {NoStop}%
\bibitem [{\citenamefont {Cao}\ \emph {et~al.}(2020)\citenamefont {Cao}, \citenamefont {Rodan-Legrain}, \citenamefont {Rubies-Bigorda}, \citenamefont {Park}, \citenamefont {Watanabe}, \citenamefont {Taniguchi},\ and\ \citenamefont {Jarillo-Herrero}}]{Cao2020}%
  \BibitemOpen
  \bibfield  {author} {\bibinfo {author} {\bibfnamefont {Y.}~\bibnamefont {Cao}}, \bibinfo {author} {\bibfnamefont {D.}~\bibnamefont {Rodan-Legrain}}, \bibinfo {author} {\bibfnamefont {O.}~\bibnamefont {Rubies-Bigorda}}, \bibinfo {author} {\bibfnamefont {J.~M.}\ \bibnamefont {Park}}, \bibinfo {author} {\bibfnamefont {K.}~\bibnamefont {Watanabe}}, \bibinfo {author} {\bibfnamefont {T.}~\bibnamefont {Taniguchi}},\ and\ \bibinfo {author} {\bibfnamefont {P.}~\bibnamefont {Jarillo-Herrero}},\ }\bibfield  {title} {\bibinfo {title} {Tunable correlated states and spin-polarized phases in twisted bilayer--bilayer graphene},\ }\href {https://doi.org/10.1038/s41586-020-2260-6} {\bibfield  {journal} {\bibinfo  {journal} {Nature}\ }\textbf {\bibinfo {volume} {583}},\ \bibinfo {pages} {215} (\bibinfo {year} {2020})}\BibitemShut {NoStop}%
\bibitem [{\citenamefont {Kazmierczak}\ \emph {et~al.}(2021)\citenamefont {Kazmierczak}, \citenamefont {Van~Winkle}, \citenamefont {Ophus}, \citenamefont {Bustillo}, \citenamefont {Carr}, \citenamefont {Brown}, \citenamefont {Ciston}, \citenamefont {Taniguchi}, \citenamefont {Watanabe},\ and\ \citenamefont {Bediako}}]{Kazmierczak2021}%
  \BibitemOpen
  \bibfield  {author} {\bibinfo {author} {\bibfnamefont {N.~P.}\ \bibnamefont {Kazmierczak}}, \bibinfo {author} {\bibfnamefont {M.}~\bibnamefont {Van~Winkle}}, \bibinfo {author} {\bibfnamefont {C.}~\bibnamefont {Ophus}}, \bibinfo {author} {\bibfnamefont {K.~C.}\ \bibnamefont {Bustillo}}, \bibinfo {author} {\bibfnamefont {S.}~\bibnamefont {Carr}}, \bibinfo {author} {\bibfnamefont {H.~G.}\ \bibnamefont {Brown}}, \bibinfo {author} {\bibfnamefont {J.}~\bibnamefont {Ciston}}, \bibinfo {author} {\bibfnamefont {T.}~\bibnamefont {Taniguchi}}, \bibinfo {author} {\bibfnamefont {K.}~\bibnamefont {Watanabe}},\ and\ \bibinfo {author} {\bibfnamefont {D.~K.}\ \bibnamefont {Bediako}},\ }\bibfield  {title} {\bibinfo {title} {Strain fields in twisted bilayer graphene},\ }\href {https://doi.org/10.1038/s41563-021-00973-w} {\bibfield  {journal} {\bibinfo  {journal} {Nature Materials}\ }\textbf {\bibinfo {volume} {20}},\ \bibinfo {pages} {956} (\bibinfo {year} {2021})}\BibitemShut {NoStop}%
\bibitem [{\citenamefont {Cao}\ \emph {et~al.}(2018{\natexlab{a}})\citenamefont {Cao}, \citenamefont {Fatemi}, \citenamefont {Fang}, \citenamefont {Watanabe}, \citenamefont {Taniguchi}, \citenamefont {Kaxiras},\ and\ \citenamefont {Jarillo-Herrero}}]{Cao2018_1}%
  \BibitemOpen
  \bibfield  {author} {\bibinfo {author} {\bibfnamefont {Y.}~\bibnamefont {Cao}}, \bibinfo {author} {\bibfnamefont {V.}~\bibnamefont {Fatemi}}, \bibinfo {author} {\bibfnamefont {S.}~\bibnamefont {Fang}}, \bibinfo {author} {\bibfnamefont {K.}~\bibnamefont {Watanabe}}, \bibinfo {author} {\bibfnamefont {T.}~\bibnamefont {Taniguchi}}, \bibinfo {author} {\bibfnamefont {E.}~\bibnamefont {Kaxiras}},\ and\ \bibinfo {author} {\bibfnamefont {P.}~\bibnamefont {Jarillo-Herrero}},\ }\bibfield  {title} {\bibinfo {title} {Unconventional superconductivity in magic-angle graphene superlattices},\ }\href {https://doi.org/10.1038/nature26160} {\bibfield  {journal} {\bibinfo  {journal} {Nature}\ }\textbf {\bibinfo {volume} {556}},\ \bibinfo {pages} {43} (\bibinfo {year} {2018}{\natexlab{a}})}\BibitemShut {NoStop}%
\bibitem [{\citenamefont {Cao}\ \emph {et~al.}(2018{\natexlab{b}})\citenamefont {Cao}, \citenamefont {Fatemi}, \citenamefont {Demir}, \citenamefont {Fang}, \citenamefont {Tomarken}, \citenamefont {Luo}, \citenamefont {Sanchez-Yamagishi}, \citenamefont {Watanabe}, \citenamefont {Taniguchi}, \citenamefont {Kaxiras}, \citenamefont {Ashoori},\ and\ \citenamefont {Jarillo-Herrero}}]{Cao2018_2}%
  \BibitemOpen
  \bibfield  {author} {\bibinfo {author} {\bibfnamefont {Y.}~\bibnamefont {Cao}}, \bibinfo {author} {\bibfnamefont {V.}~\bibnamefont {Fatemi}}, \bibinfo {author} {\bibfnamefont {A.}~\bibnamefont {Demir}}, \bibinfo {author} {\bibfnamefont {S.}~\bibnamefont {Fang}}, \bibinfo {author} {\bibfnamefont {S.~L.}\ \bibnamefont {Tomarken}}, \bibinfo {author} {\bibfnamefont {J.~Y.}\ \bibnamefont {Luo}}, \bibinfo {author} {\bibfnamefont {J.~D.}\ \bibnamefont {Sanchez-Yamagishi}}, \bibinfo {author} {\bibfnamefont {K.}~\bibnamefont {Watanabe}}, \bibinfo {author} {\bibfnamefont {T.}~\bibnamefont {Taniguchi}}, \bibinfo {author} {\bibfnamefont {E.}~\bibnamefont {Kaxiras}}, \bibinfo {author} {\bibfnamefont {R.~C.}\ \bibnamefont {Ashoori}},\ and\ \bibinfo {author} {\bibfnamefont {P.}~\bibnamefont {Jarillo-Herrero}},\ }\bibfield  {title} {\bibinfo {title} {Correlated insulator behaviour at half-filling in magic-angle graphene superlattices},\ }\href {https://doi.org/10.1038/nature26154} {\bibfield  {journal} {\bibinfo  {journal}
  {Nature}\ }\textbf {\bibinfo {volume} {556}},\ \bibinfo {pages} {80} (\bibinfo {year} {2018}{\natexlab{b}})}\BibitemShut {NoStop}%
\bibitem [{\citenamefont {Yankowitz}\ \emph {et~al.}(2019)\citenamefont {Yankowitz}, \citenamefont {Chen}, \citenamefont {Polshyn}, \citenamefont {Zhang}, \citenamefont {Watanabe}, \citenamefont {Taniguchi}, \citenamefont {Graf}, \citenamefont {Young},\ and\ \citenamefont {Dean}}]{doi:10.1126/science.aav1910}%
  \BibitemOpen
  \bibfield  {author} {\bibinfo {author} {\bibfnamefont {M.}~\bibnamefont {Yankowitz}}, \bibinfo {author} {\bibfnamefont {S.}~\bibnamefont {Chen}}, \bibinfo {author} {\bibfnamefont {H.}~\bibnamefont {Polshyn}}, \bibinfo {author} {\bibfnamefont {Y.}~\bibnamefont {Zhang}}, \bibinfo {author} {\bibfnamefont {K.}~\bibnamefont {Watanabe}}, \bibinfo {author} {\bibfnamefont {T.}~\bibnamefont {Taniguchi}}, \bibinfo {author} {\bibfnamefont {D.}~\bibnamefont {Graf}}, \bibinfo {author} {\bibfnamefont {A.~F.}\ \bibnamefont {Young}},\ and\ \bibinfo {author} {\bibfnamefont {C.~R.}\ \bibnamefont {Dean}},\ }\bibfield  {title} {\bibinfo {title} {Tuning superconductivity in twisted bilayer graphene},\ }\href {https://doi.org/10.1126/science.aav1910} {\bibfield  {journal} {\bibinfo  {journal} {Science}\ }\textbf {\bibinfo {volume} {363}},\ \bibinfo {pages} {1059} (\bibinfo {year} {2019})},\ \Eprint {https://arxiv.org/abs/https://www.science.org/doi/pdf/10.1126/science.aav1910}
  {https://www.science.org/doi/pdf/10.1126/science.aav1910} \BibitemShut {NoStop}%
\bibitem [{\citenamefont {Kerelsky}\ \emph {et~al.}(2019)\citenamefont {Kerelsky}, \citenamefont {McGilly}, \citenamefont {Kennes}, \citenamefont {Xian}, \citenamefont {Yankowitz}, \citenamefont {Chen}, \citenamefont {Watanabe}, \citenamefont {Taniguchi}, \citenamefont {Hone}, \citenamefont {Dean}, \citenamefont {Rubio},\ and\ \citenamefont {Pasupathy}}]{Kerelsky2019}%
  \BibitemOpen
  \bibfield  {author} {\bibinfo {author} {\bibfnamefont {A.}~\bibnamefont {Kerelsky}}, \bibinfo {author} {\bibfnamefont {L.~J.}\ \bibnamefont {McGilly}}, \bibinfo {author} {\bibfnamefont {D.~M.}\ \bibnamefont {Kennes}}, \bibinfo {author} {\bibfnamefont {L.}~\bibnamefont {Xian}}, \bibinfo {author} {\bibfnamefont {M.}~\bibnamefont {Yankowitz}}, \bibinfo {author} {\bibfnamefont {S.}~\bibnamefont {Chen}}, \bibinfo {author} {\bibfnamefont {K.}~\bibnamefont {Watanabe}}, \bibinfo {author} {\bibfnamefont {T.}~\bibnamefont {Taniguchi}}, \bibinfo {author} {\bibfnamefont {J.}~\bibnamefont {Hone}}, \bibinfo {author} {\bibfnamefont {C.}~\bibnamefont {Dean}}, \bibinfo {author} {\bibfnamefont {A.}~\bibnamefont {Rubio}},\ and\ \bibinfo {author} {\bibfnamefont {A.~N.}\ \bibnamefont {Pasupathy}},\ }\bibfield  {title} {\bibinfo {title} {Maximized electron interactions at the magic angle in twisted bilayer graphene},\ }\href {https://doi.org/10.1038/s41586-019-1431-9} {\bibfield  {journal} {\bibinfo  {journal} {Nature}\ }\textbf
  {\bibinfo {volume} {572}},\ \bibinfo {pages} {95} (\bibinfo {year} {2019})}\BibitemShut {NoStop}%
\bibitem [{\citenamefont {Choi}\ \emph {et~al.}(2019)\citenamefont {Choi}, \citenamefont {Kemmer}, \citenamefont {Peng}, \citenamefont {Thomson}, \citenamefont {Arora}, \citenamefont {Polski}, \citenamefont {Zhang}, \citenamefont {Ren}, \citenamefont {Alicea}, \citenamefont {Refael}, \citenamefont {von Oppen}, \citenamefont {Watanabe}, \citenamefont {Taniguchi},\ and\ \citenamefont {Nadj-Perge}}]{Choi2019_1}%
  \BibitemOpen
  \bibfield  {author} {\bibinfo {author} {\bibfnamefont {Y.}~\bibnamefont {Choi}}, \bibinfo {author} {\bibfnamefont {J.}~\bibnamefont {Kemmer}}, \bibinfo {author} {\bibfnamefont {Y.}~\bibnamefont {Peng}}, \bibinfo {author} {\bibfnamefont {A.}~\bibnamefont {Thomson}}, \bibinfo {author} {\bibfnamefont {H.}~\bibnamefont {Arora}}, \bibinfo {author} {\bibfnamefont {R.}~\bibnamefont {Polski}}, \bibinfo {author} {\bibfnamefont {Y.}~\bibnamefont {Zhang}}, \bibinfo {author} {\bibfnamefont {H.}~\bibnamefont {Ren}}, \bibinfo {author} {\bibfnamefont {J.}~\bibnamefont {Alicea}}, \bibinfo {author} {\bibfnamefont {G.}~\bibnamefont {Refael}}, \bibinfo {author} {\bibfnamefont {F.}~\bibnamefont {von Oppen}}, \bibinfo {author} {\bibfnamefont {K.}~\bibnamefont {Watanabe}}, \bibinfo {author} {\bibfnamefont {T.}~\bibnamefont {Taniguchi}},\ and\ \bibinfo {author} {\bibfnamefont {S.}~\bibnamefont {Nadj-Perge}},\ }\bibfield  {title} {\bibinfo {title} {Electronic correlations in twisted bilayer graphene near the magic angle},\ }\href
  {https://doi.org/10.1038/s41567-019-0606-5} {\bibfield  {journal} {\bibinfo  {journal} {Nature Physics}\ }\textbf {\bibinfo {volume} {15}},\ \bibinfo {pages} {1174} (\bibinfo {year} {2019})}\BibitemShut {NoStop}%
\bibitem [{\citenamefont {Serlin}\ \emph {et~al.}(2020)\citenamefont {Serlin}, \citenamefont {Tschirhart}, \citenamefont {Polshyn}, \citenamefont {Zhang}, \citenamefont {Zhu}, \citenamefont {Watanabe}, \citenamefont {Taniguchi}, \citenamefont {Balents},\ and\ \citenamefont {Young}}]{doi:10.1126/science.aay5533}%
  \BibitemOpen
  \bibfield  {author} {\bibinfo {author} {\bibfnamefont {M.}~\bibnamefont {Serlin}}, \bibinfo {author} {\bibfnamefont {C.~L.}\ \bibnamefont {Tschirhart}}, \bibinfo {author} {\bibfnamefont {H.}~\bibnamefont {Polshyn}}, \bibinfo {author} {\bibfnamefont {Y.}~\bibnamefont {Zhang}}, \bibinfo {author} {\bibfnamefont {J.}~\bibnamefont {Zhu}}, \bibinfo {author} {\bibfnamefont {K.}~\bibnamefont {Watanabe}}, \bibinfo {author} {\bibfnamefont {T.}~\bibnamefont {Taniguchi}}, \bibinfo {author} {\bibfnamefont {L.}~\bibnamefont {Balents}},\ and\ \bibinfo {author} {\bibfnamefont {A.~F.}\ \bibnamefont {Young}},\ }\bibfield  {title} {\bibinfo {title} {Intrinsic quantized anomalous hall effect in a moiré heterostructure},\ }\href {https://doi.org/10.1126/science.aay5533} {\bibfield  {journal} {\bibinfo  {journal} {Science}\ }\textbf {\bibinfo {volume} {367}},\ \bibinfo {pages} {900} (\bibinfo {year} {2020})},\ \Eprint {https://arxiv.org/abs/https://www.science.org/doi/pdf/10.1126/science.aay5533}
  {https://www.science.org/doi/pdf/10.1126/science.aay5533} \BibitemShut {NoStop}%
\bibitem [{\citenamefont {Sharpe}\ \emph {et~al.}(2019)\citenamefont {Sharpe}, \citenamefont {Fox}, \citenamefont {Barnard}, \citenamefont {Finney}, \citenamefont {Watanabe}, \citenamefont {Taniguchi}, \citenamefont {Kastner},\ and\ \citenamefont {Goldhaber-Gordon}}]{doi:10.1126/science.aaw3780}%
  \BibitemOpen
  \bibfield  {author} {\bibinfo {author} {\bibfnamefont {A.~L.}\ \bibnamefont {Sharpe}}, \bibinfo {author} {\bibfnamefont {E.~J.}\ \bibnamefont {Fox}}, \bibinfo {author} {\bibfnamefont {A.~W.}\ \bibnamefont {Barnard}}, \bibinfo {author} {\bibfnamefont {J.}~\bibnamefont {Finney}}, \bibinfo {author} {\bibfnamefont {K.}~\bibnamefont {Watanabe}}, \bibinfo {author} {\bibfnamefont {T.}~\bibnamefont {Taniguchi}}, \bibinfo {author} {\bibfnamefont {M.~A.}\ \bibnamefont {Kastner}},\ and\ \bibinfo {author} {\bibfnamefont {D.}~\bibnamefont {Goldhaber-Gordon}},\ }\bibfield  {title} {\bibinfo {title} {Emergent ferromagnetism near three-quarters filling in twisted bilayer graphene},\ }\href {https://doi.org/10.1126/science.aaw3780} {\bibfield  {journal} {\bibinfo  {journal} {Science}\ }\textbf {\bibinfo {volume} {365}},\ \bibinfo {pages} {605} (\bibinfo {year} {2019})},\ \Eprint {https://arxiv.org/abs/https://www.science.org/doi/pdf/10.1126/science.aaw3780} {https://www.science.org/doi/pdf/10.1126/science.aaw3780}
  \BibitemShut {NoStop}%
\bibitem [{\citenamefont {Tomarken}\ \emph {et~al.}(2019)\citenamefont {Tomarken}, \citenamefont {Cao}, \citenamefont {Demir}, \citenamefont {Watanabe}, \citenamefont {Taniguchi}, \citenamefont {Jarillo-Herrero},\ and\ \citenamefont {Ashoori}}]{PhysRevLett.123.046601}%
  \BibitemOpen
  \bibfield  {author} {\bibinfo {author} {\bibfnamefont {S.~L.}\ \bibnamefont {Tomarken}}, \bibinfo {author} {\bibfnamefont {Y.}~\bibnamefont {Cao}}, \bibinfo {author} {\bibfnamefont {A.}~\bibnamefont {Demir}}, \bibinfo {author} {\bibfnamefont {K.}~\bibnamefont {Watanabe}}, \bibinfo {author} {\bibfnamefont {T.}~\bibnamefont {Taniguchi}}, \bibinfo {author} {\bibfnamefont {P.}~\bibnamefont {Jarillo-Herrero}},\ and\ \bibinfo {author} {\bibfnamefont {R.~C.}\ \bibnamefont {Ashoori}},\ }\bibfield  {title} {\bibinfo {title} {Electronic compressibility of magic-angle graphene superlattices},\ }\href {https://doi.org/10.1103/PhysRevLett.123.046601} {\bibfield  {journal} {\bibinfo  {journal} {Phys. Rev. Lett.}\ }\textbf {\bibinfo {volume} {123}},\ \bibinfo {pages} {046601} (\bibinfo {year} {2019})}\BibitemShut {NoStop}%
\bibitem [{\citenamefont {Wang}\ \emph {et~al.}(2020)\citenamefont {Wang}, \citenamefont {Shih}, \citenamefont {Ghiotto}, \citenamefont {Xian}, \citenamefont {Rhodes}, \citenamefont {Tan}, \citenamefont {Claassen}, \citenamefont {Kennes}, \citenamefont {Bai}, \citenamefont {Kim}, \citenamefont {Watanabe}, \citenamefont {Taniguchi}, \citenamefont {Zhu}, \citenamefont {Hone}, \citenamefont {Rubio}, \citenamefont {Pasupathy},\ and\ \citenamefont {Dean}}]{Wang2020}%
  \BibitemOpen
  \bibfield  {author} {\bibinfo {author} {\bibfnamefont {L.}~\bibnamefont {Wang}}, \bibinfo {author} {\bibfnamefont {E.-M.}\ \bibnamefont {Shih}}, \bibinfo {author} {\bibfnamefont {A.}~\bibnamefont {Ghiotto}}, \bibinfo {author} {\bibfnamefont {L.}~\bibnamefont {Xian}}, \bibinfo {author} {\bibfnamefont {D.~A.}\ \bibnamefont {Rhodes}}, \bibinfo {author} {\bibfnamefont {C.}~\bibnamefont {Tan}}, \bibinfo {author} {\bibfnamefont {M.}~\bibnamefont {Claassen}}, \bibinfo {author} {\bibfnamefont {D.~M.}\ \bibnamefont {Kennes}}, \bibinfo {author} {\bibfnamefont {Y.}~\bibnamefont {Bai}}, \bibinfo {author} {\bibfnamefont {B.}~\bibnamefont {Kim}}, \bibinfo {author} {\bibfnamefont {K.}~\bibnamefont {Watanabe}}, \bibinfo {author} {\bibfnamefont {T.}~\bibnamefont {Taniguchi}}, \bibinfo {author} {\bibfnamefont {X.}~\bibnamefont {Zhu}}, \bibinfo {author} {\bibfnamefont {J.}~\bibnamefont {Hone}}, \bibinfo {author} {\bibfnamefont {A.}~\bibnamefont {Rubio}}, \bibinfo {author} {\bibfnamefont {A.~N.}\ \bibnamefont {Pasupathy}},\
  and\ \bibinfo {author} {\bibfnamefont {C.~R.}\ \bibnamefont {Dean}},\ }\bibfield  {title} {\bibinfo {title} {Correlated electronic phases in twisted bilayer transition metal dichalcogenides},\ }\href {https://doi.org/10.1038/s41563-020-0708-6} {\bibfield  {journal} {\bibinfo  {journal} {Nature Materials}\ }\textbf {\bibinfo {volume} {19}},\ \bibinfo {pages} {861} (\bibinfo {year} {2020})}\BibitemShut {NoStop}%
\bibitem [{\citenamefont {Xu}\ \emph {et~al.}(2020)\citenamefont {Xu}, \citenamefont {Liu}, \citenamefont {Rhodes}, \citenamefont {Watanabe}, \citenamefont {Taniguchi}, \citenamefont {Hone}, \citenamefont {Elser}, \citenamefont {Mak},\ and\ \citenamefont {Shan}}]{Xu2020}%
  \BibitemOpen
  \bibfield  {author} {\bibinfo {author} {\bibfnamefont {Y.}~\bibnamefont {Xu}}, \bibinfo {author} {\bibfnamefont {S.}~\bibnamefont {Liu}}, \bibinfo {author} {\bibfnamefont {D.~A.}\ \bibnamefont {Rhodes}}, \bibinfo {author} {\bibfnamefont {K.}~\bibnamefont {Watanabe}}, \bibinfo {author} {\bibfnamefont {T.}~\bibnamefont {Taniguchi}}, \bibinfo {author} {\bibfnamefont {J.}~\bibnamefont {Hone}}, \bibinfo {author} {\bibfnamefont {V.}~\bibnamefont {Elser}}, \bibinfo {author} {\bibfnamefont {K.~F.}\ \bibnamefont {Mak}},\ and\ \bibinfo {author} {\bibfnamefont {J.}~\bibnamefont {Shan}},\ }\bibfield  {title} {\bibinfo {title} {Correlated insulating states at fractional fillings of moir{\'e} superlattices},\ }\href {https://doi.org/10.1038/s41586-020-2868-6} {\bibfield  {journal} {\bibinfo  {journal} {Nature}\ }\textbf {\bibinfo {volume} {587}},\ \bibinfo {pages} {214} (\bibinfo {year} {2020})}\BibitemShut {NoStop}%
\bibitem [{\citenamefont {Jin}\ \emph {et~al.}(2021)\citenamefont {Jin}, \citenamefont {Tao}, \citenamefont {Li}, \citenamefont {Xu}, \citenamefont {Tang}, \citenamefont {Zhu}, \citenamefont {Liu}, \citenamefont {Watanabe}, \citenamefont {Taniguchi}, \citenamefont {Hone}, \citenamefont {Fu}, \citenamefont {Shan},\ and\ \citenamefont {Mak}}]{Jin2021}%
  \BibitemOpen
  \bibfield  {author} {\bibinfo {author} {\bibfnamefont {C.}~\bibnamefont {Jin}}, \bibinfo {author} {\bibfnamefont {Z.}~\bibnamefont {Tao}}, \bibinfo {author} {\bibfnamefont {T.}~\bibnamefont {Li}}, \bibinfo {author} {\bibfnamefont {Y.}~\bibnamefont {Xu}}, \bibinfo {author} {\bibfnamefont {Y.}~\bibnamefont {Tang}}, \bibinfo {author} {\bibfnamefont {J.}~\bibnamefont {Zhu}}, \bibinfo {author} {\bibfnamefont {S.}~\bibnamefont {Liu}}, \bibinfo {author} {\bibfnamefont {K.}~\bibnamefont {Watanabe}}, \bibinfo {author} {\bibfnamefont {T.}~\bibnamefont {Taniguchi}}, \bibinfo {author} {\bibfnamefont {J.~C.}\ \bibnamefont {Hone}}, \bibinfo {author} {\bibfnamefont {L.}~\bibnamefont {Fu}}, \bibinfo {author} {\bibfnamefont {J.}~\bibnamefont {Shan}},\ and\ \bibinfo {author} {\bibfnamefont {K.~F.}\ \bibnamefont {Mak}},\ }\bibfield  {title} {\bibinfo {title} {Stripe phases in wse2/ws2 moir{\'e} superlattices},\ }\href {https://doi.org/10.1038/s41563-021-00959-8} {\bibfield  {journal} {\bibinfo  {journal} {Nature Materials}\
  }\textbf {\bibinfo {volume} {20}},\ \bibinfo {pages} {940} (\bibinfo {year} {2021})}\BibitemShut {NoStop}%
\bibitem [{\citenamefont {Carr}\ \emph {et~al.}(2018)\citenamefont {Carr}, \citenamefont {Fang}, \citenamefont {Jarillo-Herrero},\ and\ \citenamefont {Kaxiras}}]{PhysRevB.98.085144}%
  \BibitemOpen
  \bibfield  {author} {\bibinfo {author} {\bibfnamefont {S.}~\bibnamefont {Carr}}, \bibinfo {author} {\bibfnamefont {S.}~\bibnamefont {Fang}}, \bibinfo {author} {\bibfnamefont {P.}~\bibnamefont {Jarillo-Herrero}},\ and\ \bibinfo {author} {\bibfnamefont {E.}~\bibnamefont {Kaxiras}},\ }\bibfield  {title} {\bibinfo {title} {Pressure dependence of the magic twist angle in graphene superlattices},\ }\href {https://doi.org/10.1103/PhysRevB.98.085144} {\bibfield  {journal} {\bibinfo  {journal} {Phys. Rev. B}\ }\textbf {\bibinfo {volume} {98}},\ \bibinfo {pages} {085144} (\bibinfo {year} {2018})}\BibitemShut {NoStop}%
\bibitem [{\citenamefont {Zhang}\ \emph {et~al.}(2020)\citenamefont {Zhang}, \citenamefont {Wang}, \citenamefont {Watanabe}, \citenamefont {Taniguchi}, \citenamefont {Ueno}, \citenamefont {Tutuc},\ and\ \citenamefont {LeRoy}}]{Zhang2020}%
  \BibitemOpen
  \bibfield  {author} {\bibinfo {author} {\bibfnamefont {Z.}~\bibnamefont {Zhang}}, \bibinfo {author} {\bibfnamefont {Y.}~\bibnamefont {Wang}}, \bibinfo {author} {\bibfnamefont {K.}~\bibnamefont {Watanabe}}, \bibinfo {author} {\bibfnamefont {T.}~\bibnamefont {Taniguchi}}, \bibinfo {author} {\bibfnamefont {K.}~\bibnamefont {Ueno}}, \bibinfo {author} {\bibfnamefont {E.}~\bibnamefont {Tutuc}},\ and\ \bibinfo {author} {\bibfnamefont {B.~J.}\ \bibnamefont {LeRoy}},\ }\bibfield  {title} {\bibinfo {title} {Flat bands in twisted bilayer transition metal dichalcogenides},\ }\href {https://doi.org/10.1038/s41567-020-0958-x} {\bibfield  {journal} {\bibinfo  {journal} {Nature Physics}\ }\textbf {\bibinfo {volume} {16}},\ \bibinfo {pages} {1093} (\bibinfo {year} {2020})}\BibitemShut {NoStop}%
\bibitem [{\citenamefont {Xiao}\ \emph {et~al.}(2012)\citenamefont {Xiao}, \citenamefont {Liu}, \citenamefont {Feng}, \citenamefont {Xu},\ and\ \citenamefont {Yao}}]{PhysRevLett.108.196802}%
  \BibitemOpen
  \bibfield  {author} {\bibinfo {author} {\bibfnamefont {D.}~\bibnamefont {Xiao}}, \bibinfo {author} {\bibfnamefont {G.-B.}\ \bibnamefont {Liu}}, \bibinfo {author} {\bibfnamefont {W.}~\bibnamefont {Feng}}, \bibinfo {author} {\bibfnamefont {X.}~\bibnamefont {Xu}},\ and\ \bibinfo {author} {\bibfnamefont {W.}~\bibnamefont {Yao}},\ }\bibfield  {title} {\bibinfo {title} {Coupled spin and valley physics in monolayers of ${\mathrm{mos}}_{2}$ and other group-vi dichalcogenides},\ }\href {https://doi.org/10.1103/PhysRevLett.108.196802} {\bibfield  {journal} {\bibinfo  {journal} {Phys. Rev. Lett.}\ }\textbf {\bibinfo {volume} {108}},\ \bibinfo {pages} {196802} (\bibinfo {year} {2012})}\BibitemShut {NoStop}%
\bibitem [{\citenamefont {Xia}\ \emph {et~al.}(2025)\citenamefont {Xia}, \citenamefont {Han}, \citenamefont {Watanabe}, \citenamefont {Taniguchi}, \citenamefont {Shan},\ and\ \citenamefont {Mak}}]{Xia2025}%
  \BibitemOpen
  \bibfield  {author} {\bibinfo {author} {\bibfnamefont {Y.}~\bibnamefont {Xia}}, \bibinfo {author} {\bibfnamefont {Z.}~\bibnamefont {Han}}, \bibinfo {author} {\bibfnamefont {K.}~\bibnamefont {Watanabe}}, \bibinfo {author} {\bibfnamefont {T.}~\bibnamefont {Taniguchi}}, \bibinfo {author} {\bibfnamefont {J.}~\bibnamefont {Shan}},\ and\ \bibinfo {author} {\bibfnamefont {K.~F.}\ \bibnamefont {Mak}},\ }\bibfield  {title} {\bibinfo {title} {Superconductivity in twisted bilayer wse2},\ }\href {https://doi.org/10.1038/s41586-024-08116-2} {\bibfield  {journal} {\bibinfo  {journal} {Nature}\ }\textbf {\bibinfo {volume} {637}},\ \bibinfo {pages} {833} (\bibinfo {year} {2025})}\BibitemShut {NoStop}%
\bibitem [{\citenamefont {Guo}\ \emph {et~al.}(2025)\citenamefont {Guo}, \citenamefont {Pack}, \citenamefont {Swann}, \citenamefont {Holtzman}, \citenamefont {Cothrine}, \citenamefont {Watanabe}, \citenamefont {Taniguchi}, \citenamefont {Mandrus}, \citenamefont {Barmak}, \citenamefont {Hone}, \citenamefont {Millis}, \citenamefont {Pasupathy},\ and\ \citenamefont {Dean}}]{Guo2025}%
  \BibitemOpen
  \bibfield  {author} {\bibinfo {author} {\bibfnamefont {Y.}~\bibnamefont {Guo}}, \bibinfo {author} {\bibfnamefont {J.}~\bibnamefont {Pack}}, \bibinfo {author} {\bibfnamefont {J.}~\bibnamefont {Swann}}, \bibinfo {author} {\bibfnamefont {L.}~\bibnamefont {Holtzman}}, \bibinfo {author} {\bibfnamefont {M.}~\bibnamefont {Cothrine}}, \bibinfo {author} {\bibfnamefont {K.}~\bibnamefont {Watanabe}}, \bibinfo {author} {\bibfnamefont {T.}~\bibnamefont {Taniguchi}}, \bibinfo {author} {\bibfnamefont {D.~G.}\ \bibnamefont {Mandrus}}, \bibinfo {author} {\bibfnamefont {K.}~\bibnamefont {Barmak}}, \bibinfo {author} {\bibfnamefont {J.}~\bibnamefont {Hone}}, \bibinfo {author} {\bibfnamefont {A.~J.}\ \bibnamefont {Millis}}, \bibinfo {author} {\bibfnamefont {A.}~\bibnamefont {Pasupathy}},\ and\ \bibinfo {author} {\bibfnamefont {C.~R.}\ \bibnamefont {Dean}},\ }\bibfield  {title} {\bibinfo {title} {Superconductivity in 5.0{\textdegree} twisted bilayer wse2},\ }\href {https://doi.org/10.1038/s41586-024-08381-1} {\bibfield
  {journal} {\bibinfo  {journal} {Nature}\ }\textbf {\bibinfo {volume} {637}},\ \bibinfo {pages} {839} (\bibinfo {year} {2025})}\BibitemShut {NoStop}%
\bibitem [{\citenamefont {Hsu}\ \emph {et~al.}(2021)\citenamefont {Hsu}, \citenamefont {Wu},\ and\ \citenamefont {Das~Sarma}}]{PhysRevB.104.195134}%
  \BibitemOpen
  \bibfield  {author} {\bibinfo {author} {\bibfnamefont {Y.-T.}\ \bibnamefont {Hsu}}, \bibinfo {author} {\bibfnamefont {F.}~\bibnamefont {Wu}},\ and\ \bibinfo {author} {\bibfnamefont {S.}~\bibnamefont {Das~Sarma}},\ }\bibfield  {title} {\bibinfo {title} {Spin-valley locked instabilities in moir\'e transition metal dichalcogenides with conventional and higher-order van hove singularities},\ }\href {https://doi.org/10.1103/PhysRevB.104.195134} {\bibfield  {journal} {\bibinfo  {journal} {Phys. Rev. B}\ }\textbf {\bibinfo {volume} {104}},\ \bibinfo {pages} {195134} (\bibinfo {year} {2021})}\BibitemShut {NoStop}%
\bibitem [{\citenamefont {Zegrodnik}\ and\ \citenamefont {Biborski}(2023)}]{PhysRevB.108.064506}%
  \BibitemOpen
  \bibfield  {author} {\bibinfo {author} {\bibfnamefont {M.}~\bibnamefont {Zegrodnik}}\ and\ \bibinfo {author} {\bibfnamefont {A.}~\bibnamefont {Biborski}},\ }\bibfield  {title} {\bibinfo {title} {Mixed singlet-triplet superconducting state within the moir\'e $t\text{\ensuremath{-}}j\text{\ensuremath{-}}u$ model applied to twisted bilayer ${\mathrm{wse}}_{2}$},\ }\href {https://doi.org/10.1103/PhysRevB.108.064506} {\bibfield  {journal} {\bibinfo  {journal} {Phys. Rev. B}\ }\textbf {\bibinfo {volume} {108}},\ \bibinfo {pages} {064506} (\bibinfo {year} {2023})}\BibitemShut {NoStop}%
\bibitem [{\citenamefont {Zhou}\ and\ \citenamefont {Zhang}(2023)}]{PhysRevB.108.155111}%
  \BibitemOpen
  \bibfield  {author} {\bibinfo {author} {\bibfnamefont {B.}~\bibnamefont {Zhou}}\ and\ \bibinfo {author} {\bibfnamefont {Y.-H.}\ \bibnamefont {Zhang}},\ }\bibfield  {title} {\bibinfo {title} {Chiral and nodal superconductors in the $t\text{\ensuremath{-}}j$ model with valley contrasting flux on a triangular moir\'e lattice},\ }\href {https://doi.org/10.1103/PhysRevB.108.155111} {\bibfield  {journal} {\bibinfo  {journal} {Phys. Rev. B}\ }\textbf {\bibinfo {volume} {108}},\ \bibinfo {pages} {155111} (\bibinfo {year} {2023})}\BibitemShut {NoStop}%
\bibitem [{\citenamefont {Wu}\ \emph {et~al.}(2023)\citenamefont {Wu}, \citenamefont {Wu},\ and\ \citenamefont {Yao}}]{PhysRevLett.130.126001}%
  \BibitemOpen
  \bibfield  {author} {\bibinfo {author} {\bibfnamefont {Y.-M.}\ \bibnamefont {Wu}}, \bibinfo {author} {\bibfnamefont {Z.}~\bibnamefont {Wu}},\ and\ \bibinfo {author} {\bibfnamefont {H.}~\bibnamefont {Yao}},\ }\bibfield  {title} {\bibinfo {title} {Pair-density-wave and chiral superconductivity in twisted bilayer transition metal dichalcogenides},\ }\href {https://doi.org/10.1103/PhysRevLett.130.126001} {\bibfield  {journal} {\bibinfo  {journal} {Phys. Rev. Lett.}\ }\textbf {\bibinfo {volume} {130}},\ \bibinfo {pages} {126001} (\bibinfo {year} {2023})}\BibitemShut {NoStop}%
\bibitem [{\citenamefont {Akbar}\ \emph {et~al.}(2024)\citenamefont {Akbar}, \citenamefont {Biborski}, \citenamefont {Rademaker},\ and\ \citenamefont {Zegrodnik}}]{PhysRevB.110.064516}%
  \BibitemOpen
  \bibfield  {author} {\bibinfo {author} {\bibfnamefont {W.}~\bibnamefont {Akbar}}, \bibinfo {author} {\bibfnamefont {A.}~\bibnamefont {Biborski}}, \bibinfo {author} {\bibfnamefont {L.}~\bibnamefont {Rademaker}},\ and\ \bibinfo {author} {\bibfnamefont {M.}~\bibnamefont {Zegrodnik}},\ }\bibfield  {title} {\bibinfo {title} {Topological superconductivity with mixed singlet-triplet pairing in moir\'e transition metal dichalcogenide bilayers},\ }\href {https://doi.org/10.1103/PhysRevB.110.064516} {\bibfield  {journal} {\bibinfo  {journal} {Phys. Rev. B}\ }\textbf {\bibinfo {volume} {110}},\ \bibinfo {pages} {064516} (\bibinfo {year} {2024})}\BibitemShut {NoStop}%
\bibitem [{\citenamefont {Zhu}\ \emph {et~al.}(2025)\citenamefont {Zhu}, \citenamefont {Chou}, \citenamefont {Xie},\ and\ \citenamefont {Das~Sarma}}]{PhysRevB.111.L060501}%
  \BibitemOpen
  \bibfield  {author} {\bibinfo {author} {\bibfnamefont {J.}~\bibnamefont {Zhu}}, \bibinfo {author} {\bibfnamefont {Y.-Z.}\ \bibnamefont {Chou}}, \bibinfo {author} {\bibfnamefont {M.}~\bibnamefont {Xie}},\ and\ \bibinfo {author} {\bibfnamefont {S.}~\bibnamefont {Das~Sarma}},\ }\bibfield  {title} {\bibinfo {title} {Superconductivity in twisted transition metal dichalcogenide homobilayers},\ }\href {https://doi.org/10.1103/PhysRevB.111.L060501} {\bibfield  {journal} {\bibinfo  {journal} {Phys. Rev. B}\ }\textbf {\bibinfo {volume} {111}},\ \bibinfo {pages} {L060501} (\bibinfo {year} {2025})}\BibitemShut {NoStop}%
\bibitem [{\citenamefont {Xie}\ \emph {et~al.}(2025)\citenamefont {Xie}, \citenamefont {Chen}, \citenamefont {Sur}, \citenamefont {Fang}, \citenamefont {Cano},\ and\ \citenamefont {Si}}]{PhysRevLett.134.136503}%
  \BibitemOpen
  \bibfield  {author} {\bibinfo {author} {\bibfnamefont {F.}~\bibnamefont {Xie}}, \bibinfo {author} {\bibfnamefont {L.}~\bibnamefont {Chen}}, \bibinfo {author} {\bibfnamefont {S.}~\bibnamefont {Sur}}, \bibinfo {author} {\bibfnamefont {Y.}~\bibnamefont {Fang}}, \bibinfo {author} {\bibfnamefont {J.}~\bibnamefont {Cano}},\ and\ \bibinfo {author} {\bibfnamefont {Q.}~\bibnamefont {Si}},\ }\bibfield  {title} {\bibinfo {title} {Superconductivity in twisted ${\mathrm{wse}}_{2}$ from topology-induced quantum fluctuations},\ }\href {https://doi.org/10.1103/PhysRevLett.134.136503} {\bibfield  {journal} {\bibinfo  {journal} {Phys. Rev. Lett.}\ }\textbf {\bibinfo {volume} {134}},\ \bibinfo {pages} {136503} (\bibinfo {year} {2025})}\BibitemShut {NoStop}%
\bibitem [{\citenamefont {Guerci}\ \emph {et~al.}(2024)\citenamefont {Guerci}, \citenamefont {Kaplan}, \citenamefont {Ingham}, \citenamefont {Pixley},\ and\ \citenamefont {Millis}}]{guerci2024topologicalsuperconductivityrepulsiveinteractions}%
  \BibitemOpen
  \bibfield  {author} {\bibinfo {author} {\bibfnamefont {D.}~\bibnamefont {Guerci}}, \bibinfo {author} {\bibfnamefont {D.}~\bibnamefont {Kaplan}}, \bibinfo {author} {\bibfnamefont {J.}~\bibnamefont {Ingham}}, \bibinfo {author} {\bibfnamefont {J.~H.}\ \bibnamefont {Pixley}},\ and\ \bibinfo {author} {\bibfnamefont {A.~J.}\ \bibnamefont {Millis}},\ }\href {https://arxiv.org/abs/2408.16075} {\bibinfo {title} {Topological superconductivity from repulsive interactions in twisted wse$_2$}} (\bibinfo {year} {2024}),\ \Eprint {https://arxiv.org/abs/2408.16075} {arXiv:2408.16075 [cond-mat.supr-con]} \BibitemShut {NoStop}%
\bibitem [{\citenamefont {Schrade}\ and\ \citenamefont {Fu}(2024)}]{PhysRevB.110.035143}%
  \BibitemOpen
  \bibfield  {author} {\bibinfo {author} {\bibfnamefont {C.}~\bibnamefont {Schrade}}\ and\ \bibinfo {author} {\bibfnamefont {L.}~\bibnamefont {Fu}},\ }\bibfield  {title} {\bibinfo {title} {Nematic, chiral, and topological superconductivity in twisted transition metal dichalcogenides},\ }\href {https://doi.org/10.1103/PhysRevB.110.035143} {\bibfield  {journal} {\bibinfo  {journal} {Phys. Rev. B}\ }\textbf {\bibinfo {volume} {110}},\ \bibinfo {pages} {035143} (\bibinfo {year} {2024})}\BibitemShut {NoStop}%
\bibitem [{\citenamefont {Qin}\ \emph {et~al.}(2024)\citenamefont {Qin}, \citenamefont {Qiu},\ and\ \citenamefont {Wu}}]{qin2024kohnluttingermechanismsuperconductivitytwisted}%
  \BibitemOpen
  \bibfield  {author} {\bibinfo {author} {\bibfnamefont {W.}~\bibnamefont {Qin}}, \bibinfo {author} {\bibfnamefont {W.-X.}\ \bibnamefont {Qiu}},\ and\ \bibinfo {author} {\bibfnamefont {F.}~\bibnamefont {Wu}},\ }\href {https://arxiv.org/abs/2409.16114} {\bibinfo {title} {Kohn-luttinger mechanism of superconductivity in twisted bilayer wse$_2$: Gate-tunable unconventional pairing symmetry}} (\bibinfo {year} {2024}),\ \Eprint {https://arxiv.org/abs/2409.16114} {arXiv:2409.16114 [cond-mat.supr-con]} \BibitemShut {NoStop}%
\bibitem [{\citenamefont {Chubukov}\ and\ \citenamefont {Varma}(2025)}]{PhysRevB.111.014507}%
  \BibitemOpen
  \bibfield  {author} {\bibinfo {author} {\bibfnamefont {A.~V.}\ \bibnamefont {Chubukov}}\ and\ \bibinfo {author} {\bibfnamefont {C.~M.}\ \bibnamefont {Varma}},\ }\bibfield  {title} {\bibinfo {title} {Quantum criticality and superconductivity in twisted transition metal dichalcogenides},\ }\href {https://doi.org/10.1103/PhysRevB.111.014507} {\bibfield  {journal} {\bibinfo  {journal} {Phys. Rev. B}\ }\textbf {\bibinfo {volume} {111}},\ \bibinfo {pages} {014507} (\bibinfo {year} {2025})}\BibitemShut {NoStop}%
\bibitem [{\citenamefont {Christos}\ \emph {et~al.}(2024)\citenamefont {Christos}, \citenamefont {Bonetti},\ and\ \citenamefont {Scheurer}}]{christos2024approximatesymmetriesinsulatorssuperconductivity}%
  \BibitemOpen
  \bibfield  {author} {\bibinfo {author} {\bibfnamefont {M.}~\bibnamefont {Christos}}, \bibinfo {author} {\bibfnamefont {P.~M.}\ \bibnamefont {Bonetti}},\ and\ \bibinfo {author} {\bibfnamefont {M.~S.}\ \bibnamefont {Scheurer}},\ }\href {https://arxiv.org/abs/2407.02393} {\bibinfo {title} {Approximate symmetries, insulators, and superconductivity in continuum-model description of twisted wse$_2$}} (\bibinfo {year} {2024}),\ \Eprint {https://arxiv.org/abs/2407.02393} {arXiv:2407.02393 [cond-mat.supr-con]} \BibitemShut {NoStop}%
\bibitem [{\citenamefont {Kim}\ \emph {et~al.}(2025)\citenamefont {Kim}, \citenamefont {Mendez-Valderrama}, \citenamefont {Wang},\ and\ \citenamefont {Chowdhury}}]{Kim2025}%
  \BibitemOpen
  \bibfield  {author} {\bibinfo {author} {\bibfnamefont {S.}~\bibnamefont {Kim}}, \bibinfo {author} {\bibfnamefont {J.~F.}\ \bibnamefont {Mendez-Valderrama}}, \bibinfo {author} {\bibfnamefont {X.}~\bibnamefont {Wang}},\ and\ \bibinfo {author} {\bibfnamefont {D.}~\bibnamefont {Chowdhury}},\ }\bibfield  {title} {\bibinfo {title} {Theory of correlated insulators and superconductor at $\nu$ = 1 in twisted wse2},\ }\href {https://doi.org/10.1038/s41467-025-56816-8} {\bibfield  {journal} {\bibinfo  {journal} {Nature Communications}\ }\textbf {\bibinfo {volume} {16}},\ \bibinfo {pages} {1701} (\bibinfo {year} {2025})}\BibitemShut {NoStop}%
\bibitem [{\citenamefont {Kn{\"u}ppel}\ \emph {et~al.}(2025)\citenamefont {Kn{\"u}ppel}, \citenamefont {Zhu}, \citenamefont {Xia}, \citenamefont {Xia}, \citenamefont {Han}, \citenamefont {Zeng}, \citenamefont {Watanabe}, \citenamefont {Taniguchi}, \citenamefont {Shan},\ and\ \citenamefont {Mak}}]{Knüppel2025}%
  \BibitemOpen
  \bibfield  {author} {\bibinfo {author} {\bibfnamefont {P.}~\bibnamefont {Kn{\"u}ppel}}, \bibinfo {author} {\bibfnamefont {J.}~\bibnamefont {Zhu}}, \bibinfo {author} {\bibfnamefont {Y.}~\bibnamefont {Xia}}, \bibinfo {author} {\bibfnamefont {Z.}~\bibnamefont {Xia}}, \bibinfo {author} {\bibfnamefont {Z.}~\bibnamefont {Han}}, \bibinfo {author} {\bibfnamefont {Y.}~\bibnamefont {Zeng}}, \bibinfo {author} {\bibfnamefont {K.}~\bibnamefont {Watanabe}}, \bibinfo {author} {\bibfnamefont {T.}~\bibnamefont {Taniguchi}}, \bibinfo {author} {\bibfnamefont {J.}~\bibnamefont {Shan}},\ and\ \bibinfo {author} {\bibfnamefont {K.~F.}\ \bibnamefont {Mak}},\ }\bibfield  {title} {\bibinfo {title} {Correlated states controlled by a tunable van hove singularity in moir{\'e} wse2 bilayers},\ }\href {https://doi.org/10.1038/s41467-025-57235-5} {\bibfield  {journal} {\bibinfo  {journal} {Nature Communications}\ }\textbf {\bibinfo {volume} {16}},\ \bibinfo {pages} {1959} (\bibinfo {year} {2025})}\BibitemShut {NoStop}%
\bibitem [{\citenamefont {Peng}\ \emph {et~al.}(2025)\citenamefont {Peng}, \citenamefont {Beule}, \citenamefont {Li}, \citenamefont {Yang}, \citenamefont {Mele},\ and\ \citenamefont {Adam}}]{peng2025magnetismtwistedbilayerwse2}%
  \BibitemOpen
  \bibfield  {author} {\bibinfo {author} {\bibfnamefont {L.}~\bibnamefont {Peng}}, \bibinfo {author} {\bibfnamefont {C.~D.}\ \bibnamefont {Beule}}, \bibinfo {author} {\bibfnamefont {D.}~\bibnamefont {Li}}, \bibinfo {author} {\bibfnamefont {L.}~\bibnamefont {Yang}}, \bibinfo {author} {\bibfnamefont {E.~J.}\ \bibnamefont {Mele}},\ and\ \bibinfo {author} {\bibfnamefont {S.}~\bibnamefont {Adam}},\ }\href {https://arxiv.org/abs/2503.09689} {\bibinfo {title} {Magnetism in twisted bilayer wse$_2$}} (\bibinfo {year} {2025}),\ \Eprint {https://arxiv.org/abs/2503.09689} {arXiv:2503.09689 [cond-mat.str-el]} \BibitemShut {NoStop}%
\bibitem [{\citenamefont {Tuo}\ \emph {et~al.}(2024)\citenamefont {Tuo}, \citenamefont {Li}, \citenamefont {Wu}, \citenamefont {Sun},\ and\ \citenamefont {Yao}}]{tuo2024theorytopologicalsuperconductivityantiferromagnetic}%
  \BibitemOpen
  \bibfield  {author} {\bibinfo {author} {\bibfnamefont {C.}~\bibnamefont {Tuo}}, \bibinfo {author} {\bibfnamefont {M.-R.}\ \bibnamefont {Li}}, \bibinfo {author} {\bibfnamefont {Z.}~\bibnamefont {Wu}}, \bibinfo {author} {\bibfnamefont {W.}~\bibnamefont {Sun}},\ and\ \bibinfo {author} {\bibfnamefont {H.}~\bibnamefont {Yao}},\ }\href {https://arxiv.org/abs/2409.06779} {\bibinfo {title} {Theory of topological superconductivity and antiferromagnetic correlated insulators in twisted bilayer wse${}_2$}} (\bibinfo {year} {2024}),\ \Eprint {https://arxiv.org/abs/2409.06779} {arXiv:2409.06779 [cond-mat.str-el]} \BibitemShut {NoStop}%
\bibitem [{\citenamefont {Fischer}\ \emph {et~al.}(2024)\citenamefont {Fischer}, \citenamefont {Klebl}, \citenamefont {Crépel}, \citenamefont {Ryee}, \citenamefont {Rubio}, \citenamefont {Xian}, \citenamefont {Wehling}, \citenamefont {Georges}, \citenamefont {Kennes},\ and\ \citenamefont {Millis}}]{fischer2024theoryintervalleycoherentafmorder}%
  \BibitemOpen
  \bibfield  {author} {\bibinfo {author} {\bibfnamefont {A.}~\bibnamefont {Fischer}}, \bibinfo {author} {\bibfnamefont {L.}~\bibnamefont {Klebl}}, \bibinfo {author} {\bibfnamefont {V.}~\bibnamefont {Crépel}}, \bibinfo {author} {\bibfnamefont {S.}~\bibnamefont {Ryee}}, \bibinfo {author} {\bibfnamefont {A.}~\bibnamefont {Rubio}}, \bibinfo {author} {\bibfnamefont {L.}~\bibnamefont {Xian}}, \bibinfo {author} {\bibfnamefont {T.~O.}\ \bibnamefont {Wehling}}, \bibinfo {author} {\bibfnamefont {A.}~\bibnamefont {Georges}}, \bibinfo {author} {\bibfnamefont {D.~M.}\ \bibnamefont {Kennes}},\ and\ \bibinfo {author} {\bibfnamefont {A.~J.}\ \bibnamefont {Millis}},\ }\href {https://arxiv.org/abs/2412.14296} {\bibinfo {title} {Theory of intervalley-coherent afm order and topological superconductivity in twse$_2$}} (\bibinfo {year} {2024}),\ \Eprint {https://arxiv.org/abs/2412.14296} {arXiv:2412.14296 [cond-mat.str-el]} \BibitemShut {NoStop}%
\bibitem [{\citenamefont {Devakul}\ \emph {et~al.}(2021)\citenamefont {Devakul}, \citenamefont {Cr{\'e}pel}, \citenamefont {Zhang},\ and\ \citenamefont {Fu}}]{Devakul2021}%
  \BibitemOpen
  \bibfield  {author} {\bibinfo {author} {\bibfnamefont {T.}~\bibnamefont {Devakul}}, \bibinfo {author} {\bibfnamefont {V.}~\bibnamefont {Cr{\'e}pel}}, \bibinfo {author} {\bibfnamefont {Y.}~\bibnamefont {Zhang}},\ and\ \bibinfo {author} {\bibfnamefont {L.}~\bibnamefont {Fu}},\ }\bibfield  {title} {\bibinfo {title} {Magic in twisted transition metal dichalcogenide bilayers},\ }\href {https://doi.org/10.1038/s41467-021-27042-9} {\bibfield  {journal} {\bibinfo  {journal} {Nature Communications}\ }\textbf {\bibinfo {volume} {12}},\ \bibinfo {pages} {6730} (\bibinfo {year} {2021})}\BibitemShut {NoStop}%
\bibitem [{\citenamefont {Wu}\ \emph {et~al.}(2019)\citenamefont {Wu}, \citenamefont {Lovorn}, \citenamefont {Tutuc}, \citenamefont {Martin},\ and\ \citenamefont {MacDonald}}]{PhysRevLett.122.086402}%
  \BibitemOpen
  \bibfield  {author} {\bibinfo {author} {\bibfnamefont {F.}~\bibnamefont {Wu}}, \bibinfo {author} {\bibfnamefont {T.}~\bibnamefont {Lovorn}}, \bibinfo {author} {\bibfnamefont {E.}~\bibnamefont {Tutuc}}, \bibinfo {author} {\bibfnamefont {I.}~\bibnamefont {Martin}},\ and\ \bibinfo {author} {\bibfnamefont {A.~H.}\ \bibnamefont {MacDonald}},\ }\bibfield  {title} {\bibinfo {title} {Topological insulators in twisted transition metal dichalcogenide homobilayers},\ }\href {https://doi.org/10.1103/PhysRevLett.122.086402} {\bibfield  {journal} {\bibinfo  {journal} {Phys. Rev. Lett.}\ }\textbf {\bibinfo {volume} {122}},\ \bibinfo {pages} {086402} (\bibinfo {year} {2019})}\BibitemShut {NoStop}%
\bibitem [{Note1()}]{Note1}%
  \BibitemOpen
  \bibinfo {note} {By solving the Poisson equation with double-gate boundary conditions, we obtain the $k$-dependence of $V_{ll'}(k)$ in Fig. \ref {fig:schematic}d (see Supplementary Material (SM) Section I \cite {SM_tTMD}). Since $V_{ll'}(k)$ decays fast with $k$, we will neglect the $G\protect \neq 0$ components for simplicity.}\BibitemShut {Stop}%
\bibitem [{\citenamefont {Shankar}(1994)}]{RevModPhys.66.129}%
  \BibitemOpen
  \bibfield  {author} {\bibinfo {author} {\bibfnamefont {R.}~\bibnamefont {Shankar}},\ }\bibfield  {title} {\bibinfo {title} {Renormalization-group approach to interacting fermions},\ }\href {https://doi.org/10.1103/RevModPhys.66.129} {\bibfield  {journal} {\bibinfo  {journal} {Rev. Mod. Phys.}\ }\textbf {\bibinfo {volume} {66}},\ \bibinfo {pages} {129} (\bibinfo {year} {1994})}\BibitemShut {NoStop}%
\bibitem [{\citenamefont {Furukawa}\ \emph {et~al.}(1998)\citenamefont {Furukawa}, \citenamefont {Rice},\ and\ \citenamefont {Salmhofer}}]{PhysRevLett.81.3195}%
  \BibitemOpen
  \bibfield  {author} {\bibinfo {author} {\bibfnamefont {N.}~\bibnamefont {Furukawa}}, \bibinfo {author} {\bibfnamefont {T.~M.}\ \bibnamefont {Rice}},\ and\ \bibinfo {author} {\bibfnamefont {M.}~\bibnamefont {Salmhofer}},\ }\bibfield  {title} {\bibinfo {title} {Truncation of a two-dimensional fermi surface due to quasiparticle gap formation at the saddle points},\ }\href {https://doi.org/10.1103/PhysRevLett.81.3195} {\bibfield  {journal} {\bibinfo  {journal} {Phys. Rev. Lett.}\ }\textbf {\bibinfo {volume} {81}},\ \bibinfo {pages} {3195} (\bibinfo {year} {1998})}\BibitemShut {NoStop}%
\bibitem [{\citenamefont {Hur}\ and\ \citenamefont {{Maurice Rice}}(2009)}]{HUR20091452}%
  \BibitemOpen
  \bibfield  {author} {\bibinfo {author} {\bibfnamefont {K.~L.}\ \bibnamefont {Hur}}\ and\ \bibinfo {author} {\bibfnamefont {T.}~\bibnamefont {{Maurice Rice}}},\ }\bibfield  {title} {\bibinfo {title} {Superconductivity close to the mott state: From condensed-matter systems to superfluidity in optical lattices},\ }\href {https://doi.org/https://doi.org/10.1016/j.aop.2009.02.004} {\bibfield  {journal} {\bibinfo  {journal} {Annals of Physics}\ }\textbf {\bibinfo {volume} {324}},\ \bibinfo {pages} {1452} (\bibinfo {year} {2009})},\ \bibinfo {note} {july 2009 Special Issue}\BibitemShut {NoStop}%
\bibitem [{\citenamefont {Chubukov}\ \emph {et~al.}(2008)\citenamefont {Chubukov}, \citenamefont {Efremov},\ and\ \citenamefont {Eremin}}]{PhysRevB.78.134512}%
  \BibitemOpen
  \bibfield  {author} {\bibinfo {author} {\bibfnamefont {A.~V.}\ \bibnamefont {Chubukov}}, \bibinfo {author} {\bibfnamefont {D.~V.}\ \bibnamefont {Efremov}},\ and\ \bibinfo {author} {\bibfnamefont {I.}~\bibnamefont {Eremin}},\ }\bibfield  {title} {\bibinfo {title} {Magnetism, superconductivity, and pairing symmetry in iron-based superconductors},\ }\href {https://doi.org/10.1103/PhysRevB.78.134512} {\bibfield  {journal} {\bibinfo  {journal} {Phys. Rev. B}\ }\textbf {\bibinfo {volume} {78}},\ \bibinfo {pages} {134512} (\bibinfo {year} {2008})}\BibitemShut {NoStop}%
\bibitem [{\citenamefont {Nandkishore}\ \emph {et~al.}(2012)\citenamefont {Nandkishore}, \citenamefont {Levitov},\ and\ \citenamefont {Chubukov}}]{Nandkishore2012}%
  \BibitemOpen
  \bibfield  {author} {\bibinfo {author} {\bibfnamefont {R.}~\bibnamefont {Nandkishore}}, \bibinfo {author} {\bibfnamefont {L.~S.}\ \bibnamefont {Levitov}},\ and\ \bibinfo {author} {\bibfnamefont {A.~V.}\ \bibnamefont {Chubukov}},\ }\bibfield  {title} {\bibinfo {title} {Chiral superconductivity from repulsive interactions in doped graphene},\ }\href {https://doi.org/10.1038/nphys2208} {\bibfield  {journal} {\bibinfo  {journal} {Nature Physics}\ }\textbf {\bibinfo {volume} {8}},\ \bibinfo {pages} {158} (\bibinfo {year} {2012})}\BibitemShut {NoStop}%
\bibitem [{\citenamefont {Isobe}\ \emph {et~al.}(2018)\citenamefont {Isobe}, \citenamefont {Yuan},\ and\ \citenamefont {Fu}}]{PhysRevX.8.041041}%
  \BibitemOpen
  \bibfield  {author} {\bibinfo {author} {\bibfnamefont {H.}~\bibnamefont {Isobe}}, \bibinfo {author} {\bibfnamefont {N.~F.~Q.}\ \bibnamefont {Yuan}},\ and\ \bibinfo {author} {\bibfnamefont {L.}~\bibnamefont {Fu}},\ }\bibfield  {title} {\bibinfo {title} {Unconventional superconductivity and density waves in twisted bilayer graphene},\ }\href {https://doi.org/10.1103/PhysRevX.8.041041} {\bibfield  {journal} {\bibinfo  {journal} {Phys. Rev. X}\ }\textbf {\bibinfo {volume} {8}},\ \bibinfo {pages} {041041} (\bibinfo {year} {2018})}\BibitemShut {NoStop}%
\bibitem [{\citenamefont {Lin}\ and\ \citenamefont {Nandkishore}(2019)}]{PhysRevB.100.085136}%
  \BibitemOpen
  \bibfield  {author} {\bibinfo {author} {\bibfnamefont {Y.-P.}\ \bibnamefont {Lin}}\ and\ \bibinfo {author} {\bibfnamefont {R.~M.}\ \bibnamefont {Nandkishore}},\ }\bibfield  {title} {\bibinfo {title} {Chiral twist on the high-${T}_{c}$ phase diagram in moir\'e heterostructures},\ }\href {https://doi.org/10.1103/PhysRevB.100.085136} {\bibfield  {journal} {\bibinfo  {journal} {Phys. Rev. B}\ }\textbf {\bibinfo {volume} {100}},\ \bibinfo {pages} {085136} (\bibinfo {year} {2019})}\BibitemShut {NoStop}%
\bibitem [{\citenamefont {Hsu}\ \emph {et~al.}(2020{\natexlab{a}})\citenamefont {Hsu}, \citenamefont {Wu},\ and\ \citenamefont {Das~Sarma}}]{PhysRevB.102.085103}%
  \BibitemOpen
  \bibfield  {author} {\bibinfo {author} {\bibfnamefont {Y.-T.}\ \bibnamefont {Hsu}}, \bibinfo {author} {\bibfnamefont {F.}~\bibnamefont {Wu}},\ and\ \bibinfo {author} {\bibfnamefont {S.}~\bibnamefont {Das~Sarma}},\ }\bibfield  {title} {\bibinfo {title} {Topological superconductivity, ferromagnetism, and valley-polarized phases in moir\'e systems: Renormalization group analysis for twisted double bilayer graphene},\ }\href {https://doi.org/10.1103/PhysRevB.102.085103} {\bibfield  {journal} {\bibinfo  {journal} {Phys. Rev. B}\ }\textbf {\bibinfo {volume} {102}},\ \bibinfo {pages} {085103} (\bibinfo {year} {2020}{\natexlab{a}})}\BibitemShut {NoStop}%
\bibitem [{\citenamefont {Classen}\ \emph {et~al.}(2020)\citenamefont {Classen}, \citenamefont {Chubukov}, \citenamefont {Honerkamp},\ and\ \citenamefont {Scherer}}]{PhysRevB.102.125141}%
  \BibitemOpen
  \bibfield  {author} {\bibinfo {author} {\bibfnamefont {L.}~\bibnamefont {Classen}}, \bibinfo {author} {\bibfnamefont {A.~V.}\ \bibnamefont {Chubukov}}, \bibinfo {author} {\bibfnamefont {C.}~\bibnamefont {Honerkamp}},\ and\ \bibinfo {author} {\bibfnamefont {M.~M.}\ \bibnamefont {Scherer}},\ }\bibfield  {title} {\bibinfo {title} {Competing orders at higher-order van hove points},\ }\href {https://doi.org/10.1103/PhysRevB.102.125141} {\bibfield  {journal} {\bibinfo  {journal} {Phys. Rev. B}\ }\textbf {\bibinfo {volume} {102}},\ \bibinfo {pages} {125141} (\bibinfo {year} {2020})}\BibitemShut {NoStop}%
\bibitem [{\citenamefont {Lin}\ and\ \citenamefont {Nandkishore}(2020)}]{PhysRevB.102.245122}%
  \BibitemOpen
  \bibfield  {author} {\bibinfo {author} {\bibfnamefont {Y.-P.}\ \bibnamefont {Lin}}\ and\ \bibinfo {author} {\bibfnamefont {R.~M.}\ \bibnamefont {Nandkishore}},\ }\bibfield  {title} {\bibinfo {title} {Parquet renormalization group analysis of weak-coupling instabilities with multiple high-order van hove points inside the brillouin zone},\ }\href {https://doi.org/10.1103/PhysRevB.102.245122} {\bibfield  {journal} {\bibinfo  {journal} {Phys. Rev. B}\ }\textbf {\bibinfo {volume} {102}},\ \bibinfo {pages} {245122} (\bibinfo {year} {2020})}\BibitemShut {NoStop}%
\bibitem [{\citenamefont {Lu}\ \emph {et~al.}(2022)\citenamefont {Lu}, \citenamefont {Wang}, \citenamefont {Chatterjee},\ and\ \citenamefont {You}}]{PhysRevB.106.155115}%
  \BibitemOpen
  \bibfield  {author} {\bibinfo {author} {\bibfnamefont {D.-C.}\ \bibnamefont {Lu}}, \bibinfo {author} {\bibfnamefont {T.}~\bibnamefont {Wang}}, \bibinfo {author} {\bibfnamefont {S.}~\bibnamefont {Chatterjee}},\ and\ \bibinfo {author} {\bibfnamefont {Y.-Z.}\ \bibnamefont {You}},\ }\bibfield  {title} {\bibinfo {title} {Correlated metals and unconventional superconductivity in rhombohedral trilayer graphene: A renormalization group analysis},\ }\href {https://doi.org/10.1103/PhysRevB.106.155115} {\bibfield  {journal} {\bibinfo  {journal} {Phys. Rev. B}\ }\textbf {\bibinfo {volume} {106}},\ \bibinfo {pages} {155115} (\bibinfo {year} {2022})}\BibitemShut {NoStop}%
\bibitem [{\citenamefont {Son}\ \emph {et~al.}(2025)\citenamefont {Son}, \citenamefont {Hsu},\ and\ \citenamefont {Kim}}]{PhysRevB.111.115144}%
  \BibitemOpen
  \bibfield  {author} {\bibinfo {author} {\bibfnamefont {J.~H.}\ \bibnamefont {Son}}, \bibinfo {author} {\bibfnamefont {Y.-T.}\ \bibnamefont {Hsu}},\ and\ \bibinfo {author} {\bibfnamefont {E.-A.}\ \bibnamefont {Kim}},\ }\bibfield  {title} {\bibinfo {title} {Switching between superconductivity and current density waves in bernal bilayer graphene},\ }\href {https://doi.org/10.1103/PhysRevB.111.115144} {\bibfield  {journal} {\bibinfo  {journal} {Phys. Rev. B}\ }\textbf {\bibinfo {volume} {111}},\ \bibinfo {pages} {115144} (\bibinfo {year} {2025})}\BibitemShut {NoStop}%
\bibitem [{SM_()}]{SM_tTMD}%
  \BibitemOpen
  \href@noop {} {\ }\bibinfo {note} {See Supplementary Materials at URL for details on I. Double-gate Coulomb interaction, II. RG equations.}\BibitemShut {Stop}%
\bibitem [{Note2()}]{Note2}%
  \BibitemOpen
  \bibinfo {note} {The prefactor $\nu _0$ depends on the specific dispersion obtained from Eq. \ref {eq:Moire_H}. {For the RG calculation, we take the phenomenological value $\nu _0=\protect \frac {1}{2\pi W}$ for the bandwidth $W$.}}\BibitemShut {Stop}%
\bibitem [{\citenamefont {Hsu}\ \emph {et~al.}(2017{\natexlab{a}})\citenamefont {Hsu}, \citenamefont {Vaezi}, \citenamefont {Fischer},\ and\ \citenamefont {Kim}}]{Hsu2017}%
  \BibitemOpen
  \bibfield  {author} {\bibinfo {author} {\bibfnamefont {Y.-T.}\ \bibnamefont {Hsu}}, \bibinfo {author} {\bibfnamefont {A.}~\bibnamefont {Vaezi}}, \bibinfo {author} {\bibfnamefont {M.~H.}\ \bibnamefont {Fischer}},\ and\ \bibinfo {author} {\bibfnamefont {E.-A.}\ \bibnamefont {Kim}},\ }\bibfield  {title} {\bibinfo {title} {Topological superconductivity in monolayer transition metal dichalcogenides},\ }\href {https://doi.org/10.1038/ncomms14985} {\bibfield  {journal} {\bibinfo  {journal} {Nature Communications}\ }\textbf {\bibinfo {volume} {8}},\ \bibinfo {pages} {14985} (\bibinfo {year} {2017}{\natexlab{a}})}\BibitemShut {NoStop}%
\bibitem [{\citenamefont {Hsu}\ \emph {et~al.}(2020{\natexlab{b}})\citenamefont {Hsu}, \citenamefont {Wu},\ and\ \citenamefont {Das~Sarma}}]{RG_DTBG}%
  \BibitemOpen
  \bibfield  {author} {\bibinfo {author} {\bibfnamefont {Y.-T.}\ \bibnamefont {Hsu}}, \bibinfo {author} {\bibfnamefont {F.}~\bibnamefont {Wu}},\ and\ \bibinfo {author} {\bibfnamefont {S.}~\bibnamefont {Das~Sarma}},\ }\bibfield  {title} {\bibinfo {title} {Topological superconductivity, ferromagnetism, and valley-polarized phases in moir\'e systems: Renormalization group analysis for twisted double bilayer graphene},\ }\href {https://doi.org/10.1103/PhysRevB.102.085103} {\bibfield  {journal} {\bibinfo  {journal} {Phys. Rev. B}\ }\textbf {\bibinfo {volume} {102}},\ \bibinfo {pages} {085103} (\bibinfo {year} {2020}{\natexlab{b}})}\BibitemShut {NoStop}%
\bibitem [{\citenamefont {Yang}\ and\ \citenamefont {Hsu}(2025)}]{PhysRevB.111.054514}%
  \BibitemOpen
  \bibfield  {author} {\bibinfo {author} {\bibfnamefont {H.-J.}\ \bibnamefont {Yang}}\ and\ \bibinfo {author} {\bibfnamefont {Y.-T.}\ \bibnamefont {Hsu}},\ }\bibfield  {title} {\bibinfo {title} {Optical absorption signatures of superconductors driven by van hove singularities},\ }\href {https://doi.org/10.1103/PhysRevB.111.054514} {\bibfield  {journal} {\bibinfo  {journal} {Phys. Rev. B}\ }\textbf {\bibinfo {volume} {111}},\ \bibinfo {pages} {054514} (\bibinfo {year} {2025})}\BibitemShut {NoStop}%
\bibitem [{\citenamefont {Hsu}\ \emph {et~al.}(2017{\natexlab{b}})\citenamefont {Hsu}, \citenamefont {Vaezi}, \citenamefont {Fischer},\ and\ \citenamefont {Kim}}]{Hsu_monoTMD}%
  \BibitemOpen
  \bibfield  {author} {\bibinfo {author} {\bibfnamefont {Y.-T.}\ \bibnamefont {Hsu}}, \bibinfo {author} {\bibfnamefont {A.}~\bibnamefont {Vaezi}}, \bibinfo {author} {\bibfnamefont {M.~H.}\ \bibnamefont {Fischer}},\ and\ \bibinfo {author} {\bibfnamefont {E.-A.}\ \bibnamefont {Kim}},\ }\bibfield  {title} {\bibinfo {title} {Topological superconductivity in monolayer transition metal dichalcogenides},\ }\href {https://doi.org/10.1038/ncomms14985} {\bibfield  {journal} {\bibinfo  {journal} {Nature Communications}\ }\textbf {\bibinfo {volume} {8}},\ \bibinfo {pages} {14985} (\bibinfo {year} {2017}{\natexlab{b}})}\BibitemShut {NoStop}%
\bibitem [{\citenamefont {Gao}\ \emph {et~al.}(2025)\citenamefont {Gao}, \citenamefont {Ghafariasl}, \citenamefont {Mehrabad}, \citenamefont {Huang}, \citenamefont {Zhang}, \citenamefont {Session}, \citenamefont {Upadhyay}, \citenamefont {Ma}, \citenamefont {Alshalan}, \citenamefont {Forero}, \citenamefont {Sarkar}, \citenamefont {Park}, \citenamefont {Jang}, \citenamefont {Watanabe}, \citenamefont {Taniguchi}, \citenamefont {Xie}, \citenamefont {Zhou},\ and\ \citenamefont {Hafezi}}]{gao2025probingquantumanomaloushall}%
  \BibitemOpen
  \bibfield  {author} {\bibinfo {author} {\bibfnamefont {B.}~\bibnamefont {Gao}}, \bibinfo {author} {\bibfnamefont {M.}~\bibnamefont {Ghafariasl}}, \bibinfo {author} {\bibfnamefont {M.~J.}\ \bibnamefont {Mehrabad}}, \bibinfo {author} {\bibfnamefont {T.-S.}\ \bibnamefont {Huang}}, \bibinfo {author} {\bibfnamefont {L.}~\bibnamefont {Zhang}}, \bibinfo {author} {\bibfnamefont {D.}~\bibnamefont {Session}}, \bibinfo {author} {\bibfnamefont {P.}~\bibnamefont {Upadhyay}}, \bibinfo {author} {\bibfnamefont {R.}~\bibnamefont {Ma}}, \bibinfo {author} {\bibfnamefont {G.}~\bibnamefont {Alshalan}}, \bibinfo {author} {\bibfnamefont {D.~G.~S.}\ \bibnamefont {Forero}}, \bibinfo {author} {\bibfnamefont {S.}~\bibnamefont {Sarkar}}, \bibinfo {author} {\bibfnamefont {S.}~\bibnamefont {Park}}, \bibinfo {author} {\bibfnamefont {H.}~\bibnamefont {Jang}}, \bibinfo {author} {\bibfnamefont {K.}~\bibnamefont {Watanabe}}, \bibinfo {author} {\bibfnamefont {T.}~\bibnamefont {Taniguchi}}, \bibinfo {author} {\bibfnamefont {M.}~\bibnamefont
  {Xie}}, \bibinfo {author} {\bibfnamefont {Y.}~\bibnamefont {Zhou}},\ and\ \bibinfo {author} {\bibfnamefont {M.}~\bibnamefont {Hafezi}},\ }\href {https://arxiv.org/abs/2504.11530} {\bibinfo {title} {Probing quantum anomalous hall states in twisted bilayer wse2 via attractive polaron spectroscopy}} (\bibinfo {year} {2025}),\ \Eprint {https://arxiv.org/abs/2504.11530} {arXiv:2504.11530 [cond-mat.str-el]} \BibitemShut {NoStop}%
\end{thebibliography}%
\end{document}